\documentclass[aps,11pt]{revtex4}
\usepackage{dcolumn}
\usepackage{graphicx}
\usepackage{amsmath}
\usepackage{amsfonts}
\usepackage{amssymb}
\usepackage{psfrag}
\usepackage{wrapfig}
\usepackage{subfigure}
\usepackage{makeidx}
\usepackage{bm}
\usepackage{epsf}
\usepackage{hyperref}
 \usepackage{float}
\usepackage{color}
\usepackage{cases}
\usepackage{booktabs} 
\usepackage{makecell} 
\usepackage{multirow}

\makeatletter

\newcommand{\Rmnum}[1]{\uppercase\expandafter{\romannumeral #1}}
\usepackage{amsmath} 
\begin{document} 

\title{Quasinormal modes for charged Lifshitz black holes with scalar hair}

\author{Xufen Zhang$^{1}$\footnote{yzqzxf@gmail.com}, Shan Wu$^{1}$\footnote{18155514717@163.com}, Rui-Hong Yue$^{1}$\footnote{rhyue@yzu.edu.cn}, De-Cheng Zou$^{2}$\footnote{Corresponding author: dczou@jxnu.edu.cn} \\ and Ming Zhang$^{3,4}$\footnote{Corresponding author: mingzhang0807@126.com},
}

\affiliation{
$^{1}$Center for Gravitation and Cosmology, College of Physical Science and Technology, Yangzhou University, Yangzhou 225009, China \\
$^{2}$College of Physics and Communication Electronics, Jiangxi Normal University, Nanchang 330022, China \\
$^{3}$Faculty of Science, Xihang University, Xi'an 710077 China \\
$^{4}$National Joint Engineering Research Center for Special Pump System Technology, Xihang University, Xi’an 710077, China}

\begin{abstract}
In this paper, we investigate massive charged scalar perturbations in four-dimensional charged Lifshitz-AdS black holes with scalar hair, within the framework of Einstein--Maxwell--Dilaton (EMD) gravity. Using the improved asymptotic iteration method (AIM), we compute the quasinormal modes (QNMs) and explore their dependence on key parameters, including the Lifshitz dynamical exponent $z$, the scalar field mass and charge, and the black hole charge, under various spatial curvature settings ($k=0, \pm1$). Our results reveal rich and sensitive behavior in both the real and imaginary parts of QNMs. In particular, the decay rates can exhibit monotonic or non-monotonic dependence on the black hole charge, depending on the values of $z$, $m_s$, and $q_s$. These findings highlight the significant role of field and geometric parameters in governing the dynamical stability of Lifshitz black holes and offer insights into the perturbative properties of non-AdS holographic systems.

\end{abstract}

\maketitle

\section{Introduction}\label{sec:level1}

An important extension of the holographic duality beyond high-energy physics is its application to strongly coupled systems in condensed matter physics, particularly in the study of quantum critical phenomena \cite{Kachru:2008yh}. Such systems often display an anisotropic scaling symmetry between space and time, known as Lifshitz scaling,
\begin{eqnarray}
     t\mapsto \lambda^{z}t, ~~r\mapsto \lambda^{-1}r, ~~x_{i}\mapsto \lambda x_{i},
\end{eqnarray}    
where the parameter $z \geq 1$ is referred to as the dynamical or critical exponent and quantifies the degree of anisotropy. This scaling symmetry is holographically realized by the so-called Lifshitz spacetime, described by the metric
\begin{eqnarray}
  \mathrm{d}s^{2} =- (\frac{r}{l})^{2 z} \mathrm{d} t^{2}+\frac{l^{2} }{ r^{2}}\mathrm{d} r^{2}+r^{2}\mathrm{d}^{2}\vec{x}^2,  
\end{eqnarray}
where $l$ denotes the characteristic length scale of the geometry. Lifshitz spacetimes can be regarded as anisotropic generalizations of the anti-de Sitter (AdS) spacetime, preserving a constant negative curvature while sharing many of its geometric and physical features.

The no-hair theorem\cite{Ruffini:1971bza,Bekenstein:1971hc} is a core principle in black hole physics, stating that the properties of a black hole are uniquely determined by three parameters: mass, charge, and angular momentum. This theorem has been widely supported within the frameworks of Einstein's theory of gravity and Maxwell's electromagnetic theory, as the external field of a black hole in these theories follows Gauss's law, indicating that conserved quantities are closely related to the black hole's characteristics. However, with the introduction of scalar fields and other physical fields\cite{Perry:1977wk}-\cite{Luckock:1986tr}, such as  Einstein-Yang-Mills SU(2) coupled system, black holes with dilaton hairs and black holes with Skyrme hairs, it has been found that the existence of these fields may lead to a violation of the no-hair theorem. Notably, in cases of conformally coupled scalar fields, ``hairy black holes'' can emerge, which possess additional macroscopic degrees of freedom that make their external field behavior no longer conform to the no-hair theorem\cite{Saa:1996aw,Mayo:1996mv}.
The phenomenon of spontaneous scalarization was explored in the context of neutron stars 
\cite{Harada:1998ge}, revealing that effective non-minimal coupling between the scalar field and curvature can lead to non-trivial behavior of black holes, potentially resulting in changes to the total mass. This scalarization of black holes may only manifest features under strong gravitational backgrounds, and it could provide new insights into the nature of black hole scalarization with the help of emergent technologies like gravitational wave detectors (such as LISA) and black hole shadow imaging\cite{Damour:1996ke}. David Garfinkle \textit{et al.}  first proposed a series of solutions for static, spherically symmetric charged black holes in string theory. These solutions are described by the black hole's mass, charge and the asymptotic values of   scalar field (dilaton field), with the dilaton in string theory linearly coupled to $F^2$
\cite{Garfinkle:1990qj}. And Gary W. Gibbons \textit{et al.} introduced the dilaton field to provide a unified description framework for objects such as black holes and branes in higher-dimensional theories. They also revealed the sensitive dependence of the physical properties of these objects on the dilaton coupling constant\cite{Gibbons:1987ps}. In order to address the limitations of General Relativity that become apparent under extreme conditions, researchers have explored modified gravitational theories such as String Theory, particularly the heterotic string theory, which provides higher-order corrections ranging from the Gauss-Bonnet term to non-linear electromagnetic effects\cite{Bakopoulos:2024hah}. Some other solutions are provided in \cite{Gao:2004tu,Yu:2020rqi}.
 Subsequently, Alfredo and Moreira \textit{et al.} \cite{Herrera-Aguilar:2020iti,Moreira:2023zrl,Andrade:2024cdt} have also proposed new exact black hole solutions within the framework of generalized Einstein-Maxwell-Dilaton theory, arising from the breaking of spacetime isotropic scale symmetry.

A.~Herrera-Aguilar \textit{et al.} \cite{Herrera-Aguilar:2020iti} presented generalized Einstein–Maxwell–Dilaton gravity with a nonminimal coupling $h(\phi(r))$, where the action is 
\begin{align} \label{lagrangian1}
    S & = \frac{1}{16 \pi} \int \mathrm{d}^{4} x \sqrt{-g}\Big[ \frac{R-2\Lambda}{2\kappa} -\frac{1}{2}\partial _{\mu} \phi\partial^{\mu }\phi-\frac{h\left ( \phi(r)  \right ) }{4} F_{\mu \nu }F^{\mu \nu}\Big],
\end{align}
where $g = \mathrm{det}(g_{\mu \nu})$ denotes the metric determinant, $R$ is the Ricci scalar, { }$\Lambda$ is the cosmological constant, $F_{\mu \nu}=\partial_{\mu}A_{\nu}-\partial_{\nu}A_{\mu}$ is the field strength and we recall that  $h(\phi(r))$ is a nonminimal coupling function. 
The corresponding equation of motion admit Lifshitz-type solution \cite{Herrera-Aguilar:2020iti}
\begin{eqnarray}
&&ds^{2} = -\left(\frac{r}{l}\right)^{2z} f(r) \, dt^{2} + \frac{l^{2}}{r^{2} f(r)} \, dr^{2} + \left(\frac{r}{l}\right)^{2} (dx^{2} + dy^{2}),\nonumber\\
&&f(r) =1- \frac{a}{r^{z+2}} + \frac{b}{r^{2(z+1)}}, \quad \phi(r) = \sqrt{\frac{2(z-1)}{\kappa}} \, \ln\left(\frac{r}{p}\right) ,
\end{eqnarray}
where $a$ and $b$ are integration constants with dimensions of length, $p$ is the characteristic length scale, and $z \ge 1$ to ensure the reality of the scalar field. These solutions represent asymptotically Lifshitz black holes supported by both electromagnetic and scalar fields.

  
Gravitational wave detection has made significant advances in astrophysics and the study of black holes, providing a reliable method for investigating black hole oscillations or mergers\cite{Konoplya:2011qq,Kokkotas:1999bd,Wang:2005vs,Berti:2007dg}. Modern detectors like LIGO and Virgo effectively capture these signals, and by analyzing the frequency and amplitude in the waveforms, one can infer the mass, spin, and merger process of black holes. This deepens our understanding of black holes and offers new avenues for testing fundamental gravitational theories.  Black holes interact with surrounding matter, producing gravitational waves that manifest as brief radiation bursts and damped oscillations, known as quasinormal modes (QNMs). QNMs are crucial for describing dissipative systems and black hole physics, the real part of the complex frequency represents the oscillation frequency, and the imaginary part reflects the decay time scale.  Analyzing the QNMs from astrophysical black holes helps in understanding the existence of extra dimensions in string theory\cite{Chakraborty:2017qve,Aneesh:2018hlp,Visinelli:2017bny} and providing evidence for echoes in the ring-down signal\cite{Cardoso:2019rvt}.
In addition, QNMs helps to infer the mass and angular momentum of the black hole and test the no-hair theorem  of general relativity\cite{Berti:2005ys,Isi:2019aib}. QNMs play an important role in assessing the stability of background spacetimes and are key to identifying the existence of black holes and their dynamic stability. This paper will investigate the QNMs of the charged Lifshitz black holes with scalar hair, exploring the impact of model parameters on the dynamics of perturbation waves and testing the stability of the background configuration.

The organization of this paper is as follows: In Sec.~\ref{sec:level2}, we introduce the theoretical framework of $D$ dimensional Lifshitz black holes with non-minimal coupling between scalar and Maxwell fields, and discuss scalarized Lifshitz black holes for various horizon topologies $(k = 0, \pm 1)$. In Sect.~\ref{sec:level3}, we investigate massive charged scalar perturbations on the four-dimensional case.
In Sect.~\ref{sec:level4}, we use an improved asymptotic iteration method (AIM) to calculate the QNM frequencies of charged four dimensional Lifshitz black holes. In Sect.~\ref{sec:level5},  we analyze the numerical results of the QNMs and study the effects of factors such as the curvature, mass, and critical dynamical exponent of the black hole on the QNMs. Finally, we conclude the paper with closing remarks in Sec.~\ref{sec:level6}. Note that we adopt the metric signature $(-, +, +, +,\cdots)$ for higher dimensional spacetime and $(-, +, +, +)$ for 4 dimensional spacetime.

\section{Charged Lifshitz Solutions in Higher Dimensional Spacetime}\label{sec:level2}

In this study, we investigate the higher dimensional solutions in generalized Einstein-Maxwell-Dilaton theory. The action of our model is given by
\begin{align} \label{lagrangian}
    S & = \frac{1}{16 \pi} \int \mathrm{d}^{D} x \sqrt{-g}\Big[\frac{R-2\Lambda}{2\kappa} -\frac{1}{2}\partial _{\mu} \phi\partial^{\mu }\phi-\frac{h\left ( \phi (r) \right ) }{4} F_{\mu \nu }F^{\mu \nu}\Big],
\end{align}
where $g = \mathrm{det}(g_{\mu \nu})$ denotes the metric determinant,{ }$R$ is the Ricci scalar, { }$\Lambda$ is the cosmological constant, $F_{\mu \nu}=\partial_{\mu}A_{\nu}-\partial_{\nu}A_{\mu}$ is the standard Maxwell field and we recall that  $h(\phi(r))$ is a nonminimal coupling function.

Varying the action \eqref{lagrangian} with respect to the metric $g_{\mu\nu}$, dilaton field $\phi$, and electromagnetic potential $A_{\mu} $ leads to
\begin{align} 
    G_{\mu\nu}+\Lambda g_{\mu\nu } -\kappa T_{\mu\nu}=0; \label{Einstein-equ1}\\
    \nabla^2 \phi-\frac{1}{4}\frac{\mathrm{d} h}{\mathrm{d} \phi }F _{\mu\nu  } F^{\mu\nu } =0;\label{scalar-equ1}\\
    \nabla_\nu (h(\phi(r)) F^{\mu\nu})=0,\label{vector-equ1}
 \end{align}
where $G_{\mu\nu}=R_{\mu\nu}-\frac{1}{2} Rg_{\mu\nu}$ is the Einstein tensor, $T_{\mu\nu}$ is the energy-momentum tensor
 \begin{eqnarray}
  T_{\mu\nu}=\partial_{\mu}\phi \partial _{\nu}\phi-\frac{1}{2}g_{\mu\nu}
  (\partial\phi  )^2+F_{\mu \alpha }F_{\nu}{ }^{\alpha }-\frac{1}{4} F_{\alpha \beta }{ }^{\alpha \beta } g_{\mu\nu}.
 \end{eqnarray}
 
We assume the metric of Lifshitz spacetime taking the following form
\begin{eqnarray} \label{SD}
  \mathrm{d}s^{2} =-\left (\frac{r}{l}   \right ) ^{2z}f\left ( r \right )  \mathrm{d} t^{2}+\frac{ \mathrm{d} r^{2}}{\left ( \frac{r}{l}  \right )^{2}f\left ( r \right )  } + r^{2}\sum_{k=1}^{D-2} \mathrm{d}^{2}\Omega_{k},
\end{eqnarray}
where $z$ is dynamical critical exponent and $\mathrm{d}^{2} \Omega_k$  designates the line element of an $(D-2)$-dimensional hypersurface with constant curvature $(D-2)(D-3)k$. It is given by
\begin{align}
\mathrm{d}^{2} \Omega_k & =\left\{
\begin{aligned} 
&\mathrm{d}\theta_1^{2}+\sin^{2}\theta_1 (\mathrm{d}\theta_2^{2} +\sin^2{\theta_2}( +\dots + \sin^2{\theta_{D-3}}\mathrm{d}^2\theta_{D-2})) \quad\quad\quad&&k=1;\\
&\mathrm{d}\theta_1^{2}+\mathrm{d}\theta_2^{2}+\dots \mathrm{d}\theta_{D-2}^{2} \quad\quad\quad   \quad\quad\quad\quad\quad\quad  \quad\quad\quad\quad\quad\quad\quad\quad          &&k=0;\\
&\mathrm{d}\theta_1^{2}+\sinh^{2}\theta_1 (\mathrm{d}\theta_2^{2} +\sinh^2{\theta_2}( +\dots + \sinh^2{\theta_{D-3}}\mathrm{d}^2\theta_{D-2}))  \quad      &&k=-1.\\
\end{aligned}
\right .
\end{align}
 where $D$ represents the total dimension of the spacetime. This hypersurface metric can take different forms depending on the curvature (i.e., the topology), which can be spherical, flat, or hyperbolic in nature. Specifically, for $k = 1$, the curvature corresponds to positive curvature, representing spherical geometry; for $k = 0$, the curvature is zero, corresponding to flat  geometry; and for $k = -1$, the curvature is negative, corresponding to hyperbolic geometry. Each value of $k$ defines a distinct geometric structure for the angular part of the metric, which in turn influences the overall topology and physical properties of the spacetime.

Note that the fields to inherit the spacetime isometries such that they are functions of the $r$ coordinate only. The vector potential is considered to be of a purely electrical form, therefore, the field ansatzes are
\begin{align}\label{Ar}
  \phi=\phi(r), \quad A=A_t(r)\mathrm{d}t.
\end{align}

Substituting the metric ansatz (\ref{SD}) and Eq. \eqref{Ar} into the Maxwell equation (\ref{vector-equ1}), we  have
\begin{align}
{\partial _r}(\sqrt{-g}h(\phi(r))F^{rt})=0\label{V First derivative},
\end{align}
which leads to
\begin{align} \label{vector-equ2}
  {h}'(\phi(r)){A_t}'(r){\phi (r)}'r+h(\phi(r))\Big[-(z+1-D){A_t}'(r)+r{A_t}''(r)\Big] =0.
\end{align}  
We can also rewritten it as 
\begin{align} \label{vector-equ3}
\Big[ r^{D-1-z}{A_t}'(r)h(\phi(r))\Big]' =0.
\end{align} 
The integration of Eq.(\ref{vector-equ3}) gives
\begin{align}
{A_t}'(r)=\frac{{r} ^{-(D-1-z)}}{h(\phi(r))} C_0,
\end{align}
with $C_0$ representing the  constant. Let $C_0 = Ql^{D-1-z}$,  then the above equation can be written as
\begin{align}\label{atdr}
{A_t}'(r)=\frac{Q}{h(\phi(r))} \left(\frac{l}{r} \right)^{D-1-z},
\end{align}
in which $Q$ is an arbitrary constant, and takes the role of the electric charge.

Dealing with $tt$ and $rr$ components of Einstein field equations (\ref{Einstein-equ1}), one can find the identity
 \begin{eqnarray}
 E^t{}_t- E^r{}_{r}&=&\frac{ f(r)[(D-2)(1-z)+r^{2} \kappa  {\phi}' (r)^2]}{l^2} =0,\label{Ett-Err}
 \end{eqnarray} 
which leads to
 \begin{eqnarray} 
 \phi(r)=\sqrt{\frac{(D-2)(z-1)}{\kappa } } \ln(\frac{r}{p}) .\label{phir}
 \end{eqnarray}
where $p$ is a constant with the dimension of length, and $z \geq 1$ is required to ensure the reality of the scalar field.
We further consider the components of Einstein field equations
 \begin{eqnarray}
 E^r{}_{r}+ E^{\theta_1}{}_{\theta_1}&=&2\Lambda-\frac{k(D-3)^2}{r^2} +\frac{(z+D-3)(z+D-2)f(r)}{l^2} \nonumber\\
 &&+\frac{r(3(z+D-2)-1){f}'(r)}{2l^2}+\frac{r^2{f}''(r) }{2l^2}=0.\label{Err+Etheta}
 \end{eqnarray} 
Then, we obtain the solution $f(r)$
 \begin{eqnarray} 
f(r)=1+\frac{kl^2(D-3)^2}{r^2(z+D-4)^2}-\frac{a}{r^{ z+D-2}}+\frac{b}{r^{2(z+D-3)}}, \label{fr}
 \end{eqnarray}
where $a$ and $b$ are integration constants, and the parameter $ \Lambda$ euqals to
 \begin{eqnarray} \label{lambdarD}
  \Lambda =-\frac{(z+D-3)(z+D-2)}{2l^2}\label{lamb} ,
\end{eqnarray} 
to satisfy the asymptotic condition $\lim_{r \to \infty} g=g_{Lifshitz}$. 

Substituting the solutions ${A_t}'(r)$  \eqref{atdr} into $tt$ component of equation of motion
 \begin{eqnarray} 
 E^t{}_{t}&=&\frac{1}{2l^2}\Big ( 2l^2\Lambda+(D-2)r f'(r) +r^2(\frac{r}{l})^{-2z}\kappa h(\phi(r))A_t'(r)^2\nonumber\\
&& +f(r)((D-2)(D-1)+r^2\kappa \phi'(r)^2)\Big )-\frac{k(D-2)(D-3)}{2r^2}=0,\label{Ett}
\end{eqnarray}
we obtain the coupling function $h(\phi(r))$
 \begin{eqnarray}\label{hphi0}
h(\phi(r))&=&-\Big(Q^2l^Dr^{3-D}\kappa\Big)\Big[l^2(2\Lambda-kr^{-2}(D-2)(D-3))\nonumber\\
&&+(D-2)rf'(r)+f(r)(D-1)(D-2)+\kappa\phi'(r)^2\Big]^{-1}.
 \end{eqnarray}
From Eqs.(\ref{phir})(\ref{fr})(\ref{lamb}), Eq.\eqref{hphi0} becomes
 \begin{eqnarray}
h(\phi(r))&=&Q^2l^{D}\kappa(D-4+z) \Big[br^{3-z}(D-2)(D+z-4)^2+
r^{D-3}(z-1)\nonumber\\
&&\times\Big((D-3)(D-2)kl^2+D^2+z^2+2D(z-3)-6z+8\Big)\Big]^{-1}\label{hphi}.
  \end{eqnarray}
  
Considering the scalar field \eqref{phir}, we present $r$ with scalar field $\phi$
\begin{eqnarray} 
r(\phi)=pe^{\sqrt{\frac{\kappa}{(D-2)(z-1)}}\phi}\label{1phi}.
 \end{eqnarray}
Therefore, we can obtain the coupling finction $h(\phi)$ from Eqs. (\ref{hphi})(\ref{1phi})
\begin{eqnarray} \label{hphiD}
     h(\phi)& =&Q^2\kappa(D+z-4)\Big[(z-1)(D^2-5D+6)kl^{2(3-D)}p^{2(D-3)}e^{2(D-z-3)\sqrt{\frac{\kappa}{(D-2)(z-1)}}\phi}\nonumber\\
&&+(D-2)(D+z-4)^2bl^{2(2-D)}p^{2(1-z)}e^{2(1-2z)\sqrt{\frac{\kappa}{(D-2)(z-1)}}\phi}\nonumber\\
&&+(z-1)((D+z)^2-6(D+z)+8)l^{4-2D}p^{2D-4}e^{2(D-z-2)\sqrt{\frac{\kappa}{(D-2)(z-1)}}\phi}\Big]^{-1}.
\end{eqnarray}
Suppose Maxwell equations accept trivial first integrals, we integrate equation (\ref{V First derivative}) and the gauge potential $A_t(r)$ is
\begin{eqnarray}\label{Vr}
    A_t(r)&=&A_{t0}-\frac{l^{1-z-D} r^{ z+D-4} }{\kappa  Q  (D+z-4)^2}\Big(
   b (D-2) (D+z-4)^2 r^{-2 (D+z-4)}\nonumber\\
   &&-(D-3) (D-2) k l^2 (z-1)-r^2 (z-1)
   (D+z-4)^2\Big).
\end{eqnarray}
where $A_{t0}$ is an integration constant.


To confirm the geometry remains nonsingular at infinity, we examine local curvature invariants Ricci scalar $R$, Ricci square $R_{\mu\nu}R^{\mu\nu}$ and
Kretschmann scalar $R_{\mu\nu\rho\sigma}R^{\mu\nu\rho\sigma}$, which are calculated as
\begin{eqnarray}
  &&\lim_{r\to \infty}R=\frac{4 \Lambda ^2 \left(z^2+2 z+3\right)}{(z+1) (z+2)} ;\nonumber\\
  &&\lim_{r\to \infty}R_{\mu\nu}R^{\mu\nu}=\frac{8 \Lambda ^4 \left(z^4+2 z^3+5 z^2+4 z+6\right)}{(z+1)^2 (z+2)^2}; \nonumber\\
  &&\lim_{r\to \infty}R_{\mu\nu\rho\sigma}R^{\mu\nu\rho\sigma}=\frac{16 \Lambda ^4 \left(z^4+2 z^2+3\right)}{(z+1)^2 (z+2)^2} .
\end{eqnarray}
Evidently, although the scalar field does not approach a constant, all local curvature invariants approach their Lifshitz values at infinity, and the geometry is asymptotically Lifshitz in the sense of curvature invariants. The horizon is regular provided 
$f(r)$ has a simple zero.

From Eq.(\ref{scalar-equ1}), the scalar field equation reads as
 \begin{equation} \label{scalar-equ3}
  \frac{r^2\left ( \frac{r}{l}  \right )^{-2z}{h}'(\phi){A_t}' (r)^2  }{2l^2} +\frac{r(r{f}'(r){\phi }'(r)+f(r)((z+D-1){\phi }'(r)+r{\phi }''(r)))}{l^2} =0.
\end{equation}
Substituting Eqs. \eqref{phir},  \eqref{fr}, \eqref{hphiD} and  \eqref{Vr} into above equation, we find this equation \eqref{scalar-equ3} is satisfied.


\section{Massive Charged Scalar Field Perturbations}\label{sec:level3}

In this section, we focus primarily on the four-dimensional case. This choice is motivated by the physical relevance of (3+1)-dimensional spacetimes: our observable universe, where gravitational waves have been detected, is four-dimensional. This setting also allows for a cleaner analysis of quasinormal mode structures and scalar field perturbations.
The solution can be degenerated from Eq.(\ref{fr}):
\begin{eqnarray} 
 f(r)=1 + \frac{kl^2}{r^2z^2}-\frac{a}{r^{ z+2}}+\frac{b}{r^{2(z+1)}}, \label{fr1} 
 \end{eqnarray}
Moreover, the coupling function $h(\phi)$ in four dimensional spacetime can be obtained from Eq.(\ref{hphiD})
\begin{eqnarray}
h(\phi)&=&l^4Q^2z\kappa\Big[2kl^2p^2(z-1)e^{\sqrt{2}(1-z)\sqrt{\frac{\kappa}{z-1}}\phi}+2bz^2p^{2-2z}e^{\sqrt{2}(1-2z)\sqrt{\frac{\kappa}{z-1}}\phi}\nonumber\\
&&+zp^4(z^2+z-2)e^{\sqrt{2}(2-z)\sqrt{\frac{\kappa}{z-1}}\phi}\Big]^{-1}. \label{hphi1}
\end{eqnarray}
For $z=1$, Eq.\eqref{fr1} reduces to RN-AdS black hole solution and $\phi(r)$ equals to 0 from Eq.\eqref{phir}.
Moreover, the function $h(0)$ becomes $\frac{Q^2 l^6\kappa}{2b}$ from Eq.(\ref{hphi1}). With regard to the action \eqref{lagrangian}, we can further set \(h(0) = 1\) to obtain the reduced action corresponding RN-AdS black hole solution. It implies that the parameter $b$ equals to \(\frac{Q^2 l^6\kappa}{2}\) from  $h(0) = \frac{Q^2 l^6\kappa}{2b}=1$.
We set $\kappa=1$ for simplification in future.

With regard to the charged Lifshitz black hole with scalar hair (Eq.\eqref{fr1}), the parameter $b$ is related with the electric charge $Q$, scalar field parameter $p$ and scalar field $\phi(r)$ from Eq.\eqref{hphi1}. In particular, Refs.\cite{Becar:2015kpa,KordZangeneh:2017zgg,KordZangeneh:2015yyc} adopted the dilaton coupling function $h(\phi)=e^{-2\zeta \phi}$ to obtain charged lifshitz black holes with scalar hair, and then
the parameter $b$ can be expressed by parameters $p$ and $Q$ from Eq.(\ref{hphi1}). Actually, the phenomenon also appears in the Einstein–Maxwell–Scalar theory \cite{Fernandes:2019rez,Myung:2019oua,Myung:2018jvi},  where the parameter $b$ refers to the combination of electric charge $Q$ and scalar field parameter $Q_s$ for charged scalarized black hole. 

Now we investigate a massive charged scalar field perturbation on this Lifshitz black hole. The dynamical wave equation is 
\begin{eqnarray}
D^{\nu }D_{\nu } \psi = m_s^2\psi,
\end{eqnarray}
where $D^{\nu } =\nabla ^{\nu }-iq_{s} A^{\nu }$ is the gauge covariant derivative. Here $m_{s}$ is the mass of the scalar field $\psi$ and $q_{s}$ the test charge of the scalar field. 

Now we decompose $\psi$ into the following standard form:
\begin{eqnarray}
    \psi(t,r,angles)= e^{-i\omega t}R(r)Y(angles),
\end{eqnarray}
where $\omega$ is the frequency and $Y(angles)$ is the spherical harmonic function related
to the angular coordinates. Since $Y$ is the spherical harmonic function, it satisfies the following equation
\begin{eqnarray}
    \nabla _{angles}^{2} Y(\theta ,\phi )=-L(L+1)Y(angles),
\end{eqnarray}
where $L(L+1)$ is the constant of separation. 

In addition, the differential equation of the radial function in the four dimensional Lifshitz-dilaton background (\ref{SD}) is 
\begin{eqnarray}\label{diff1}
    f{R}'' +({f}' +\frac{(z+3)f}{r}){R}' + (\frac{q_{s}A_{t}+\omega }{r^{z+1} } )^{2}\frac{R}{f} -(m_s^2+\frac{L(L+1)}{r^{2} } )\frac{R}{r^{2} }=0 . 
\end{eqnarray}
Here the gauge potential ${A_t}(r)$ is given by \cite{Wu:2024aaf}

\begin{eqnarray}\label{V}
{A_t}(r)&=&A_{t0}-\frac{2 l^2  r^zz^{-2}(z-1)-z^2 \left(2 br^{-z}-(z-1) r^{z+2}\right)}{\kappa  Ql^{z+3} }.
    \end{eqnarray}


Therefore, introducing a new radial function $R(r)$ as $R(r) =
K(r)/r$ and using a tortoise coordinate $r_{*} $ as $dr_{*}/dr=[r^{z+1}f(r)]^{-1}$, the Eq. (\ref{diff1}) can be transformed into an equation resembling the Schrödinger equation
\begin{eqnarray}\label{schr}
\frac{d^{2}K(r_{*})  }{dr_{*}^{2} } +[(\omega +q_{s}A_{t})^2-V_s(r)]K(r_{*})=0 ,   
\end{eqnarray}
where the potential $V_{s}(r)$ is given by
\begin{eqnarray}
    V_{s}(r)=(z+1)f^2r^{2z}+fr^{2z}(m_s^2+\frac{L(L+1)}{r^2}) +f{f}'r^{2z+1} .
\end{eqnarray}
To achieve a comprehensive understanding of the system's behavior, we need to conduct a numerical analysis. 

\section{QUASINORMAL MODE FREQUENCIES}\label{sec:level4}

Computing the quasinormal mode (QNM) spectrum reduces to solving a nontrivial eigenvalue problem. Several numerical approaches have been developed to tackle this challenge, including highly accurate techniques such as the pseudospectral method. In this section, we employ the Improved Asymptotic Iteration Method (AIM), a powerful semi-analytical method tailored for second-order differential equations. AIM constructs a termination condition through an iterative procedure, enabling efficient and accurate extraction of QNM frequencies, particularly in spacetimes with intricate boundary behavior.

The key idea of the improved asymptotic iteration method (AIM) is to transform the second-order linear differential equation into a recursive form and iteratively extract a quantization condition that yields the eigenvalues. Specifically, we consider a differential equation of the form
\begin{eqnarray}
\chi''(x) = \zeta_0(x) \chi'(x) + s_0(x) \chi(x),
\end{eqnarray}
where $\zeta_0(x)$ and $s_0(x)$ are smooth functions of the independent variable.Through repeated differentiation, the equation generates a sequence of functions $\zeta_n(x)$ and $s_n(x)$, defined by the recurrence relations:
\begin{eqnarray}
  \zeta_{n}(x) = \zeta'_{n-1}(x) + s_{n-1}(x) + \zeta_0(x)\zeta_{n-1}(x), \quad
s_{n}(x) = s'_{n-1}(x) + s_0(x)\zeta_{n-1}(x).
\end{eqnarray}
The method terminates when the quantization condition
\begin{eqnarray}
  \delta_n(x) = s_n(x)\zeta_{n-1}(x) - s_{n-1}(x)\zeta_n(x) = 0,
\end{eqnarray}
is satisfied at some fixed point $x = x_0$. Solving this condition yields the eigenvalues, such as the quasinormal frequencies in black hole spacetimes.

To apply the AIM framework to our problem, we first rewrite the radial equation in a more suitable form. This is achieved by introducing the new dimensionless variable $u = 1 - \frac{r_h}{r}$, which maps the domain $r \in [r_h, \infty)$ to $u \in [0, 1)$, with the black hole horizon located at $u = 0$ and the asymptotic boundary at $u = 1$. 

Under this transformation $u=1-\frac{r_h}{r} $, the radial equation Eq. (\ref{diff1}) becomes
\begin{eqnarray}\label{bian}
  \frac{r_{h}^{z+1} f(u)}{(1-u)^{z-1}} R^{\prime \prime}(u) +\frac{r_{h}^{z+1}}{(1-u)^{z}}\left[(z+1) f(u)+(1-u) f^{\prime}(u)\right] R^{\prime}(u)  +\frac{r_{h}^{1-z} l^{2 z+2}}{f(u)(1-u)^{1-z}}\nonumber\\
  \left(q_{s} A_{t}(u)+\omega\right)^{2} R(u) -\frac{r_{h}^{z+1}}{(1-u)^{z+1}}\left(m_{s}^{2}+\frac{L(L+1)}{r_{h}^{2}}(1-u)^{2}\right) R(u)=0.
\end{eqnarray}
In order to propose an ansatz for Eq.(\ref{bian}), we are going to consider the behavior of the function $R(u)$ on horizon $ (u = 0)$
and boundary $(u = 1)$. At horizon $(u = 0)$, we have $f(0)\approx u{f}'(0) $ and $A_{t}(0)=0$, thus Eq.(\ref{bian}) can be written as
\begin{eqnarray}
    R^{\prime \prime}(u)+\frac{R^{\prime}(u)}{u}+\frac{R(u)}{r_{h}^{2 z} u^{2} f^{\prime 2}} \omega^{2}=0.
\end{eqnarray}
The solution for this equation can be written as
\begin{eqnarray}
    R(u\to 0)\sim C_{1}u^{-\xi }+C_{2}u^{\xi }, ~~\xi=\frac{i\omega }{r_{h}^{z} {f}'(0) } . 
\end{eqnarray}
Here we have imposed the ingoing boundary condition at the
horizon $(u = 0)$, and so we have required $C_{2}$ to vanish.
At infinity, where $(u = 1)$, Eq. (\ref{bian}), can be written as
\begin{eqnarray}
    {R}'' (u)+\frac{(z+1){R}'(u)}{1-u}-\frac{m_{s}^2R(u)}{(1-u)^2}=0. 
\end{eqnarray}
The solution to this equation can be written as
\begin{eqnarray}
    R(u\to1)\sim D_{1}(1-u)^{\frac{1}{2}(z+2+\Pi )}+D_{2}(1-u)^{\frac{1}{2}(z+2-\Pi)},
\end{eqnarray}
where 
\begin{eqnarray}
    \Pi=\sqrt{\left ( z+2 \right )^2+4m_{s}^2 } .
\end{eqnarray}
In order to impose Dirichlet boundary condition $R(u\to
1)\to  0$, we can set $D_2 = 0$.
Using the above solutions at horizon and boundary, the
general ansatz for Eq. (\ref{bian}), can be written as
\begin{eqnarray}\label{hebin}
    R(u)=u^{-\xi }(1-u)^{\frac{1}{2}(z+2+\Pi )}\chi (u).
\end{eqnarray}
Inserting Eq. (\ref{hebin}) into (\ref{bian}), we obtain
\begin{eqnarray}\label{chi}
    {\chi }'' =\zeta _0(u){\chi }' +s_0(u)\chi ,
\end{eqnarray}
where the coefficient functions are given by
\begin{eqnarray}
    \zeta _0=\frac{2i\omega }{r_h^z u{f}'(0) } -\frac{ {f}'(u) }{f(u)}+\frac{\Pi +1}{1-u} .
\end{eqnarray}
Furthermore, $s_0$ is given by
\begin{eqnarray}
s_0(u)&=&\frac{r_h^{-2(z+1)}}{2(u-1)^2u^2f(u)^2{f}'^2}\times\left[-2r_h^2u^2 f'^2(1-u)^{2z}(q_s A_t(u)+\omega)^2 \right.\nonumber\\
&+&\left.u f(u){f}'(0)r_h^z(2u{f}'(0)r_h^z(m_s^2r_h^2+L(L+1)(u-1)^2)+r_h^2(u-1){f}'(u)\right.\nonumber\\
&\times&\left.(-u{f}'(0)r_h^z(\Pi +z+2)+2i(u-1)\omega)\right.\nonumber\\
&+&\left.2r_h^2f(u)^2(i(u-1)\omega{f}'(0)r_h^z(u\Pi +1)-m_s^2u^2{f}'^2r_h^{2z}+(u-1)^2\omega^2 )\right].
\end{eqnarray}

Now we can employ  the improved asymptotic iteration method (AIM) 
to numerically solve the Eq. (\ref{chi}). In the following, we will  set $L=0$ and $l=p=1$, and analyze scalar perturbations and explore the impact of different model parameters on the real and imaginary components of the quasinormal frequencies related to Lifshitz black hole solutions.

\section{NUMERICAL RESULTS}
\label{sec:level5}

In this section, we report our numerical results of the QNMs for the scalar field around the Lifshitz black hole. We will consider the four dimensional Lifshitz model under three different geometric properties ($k=0, \pm 1$). The frequencies of the quasinormal modes are divided into real and imaginary parts, which determine the energy of the scalar perturbations and the stability of the system under dynamical perturbations.

\subsection{Massless Charged Scalar Perturbation}
\begin{figure}[H]
\centering 
\subfigure[$z=1$]{\label{sfm0kza} 
\includegraphics[width=1.2in]{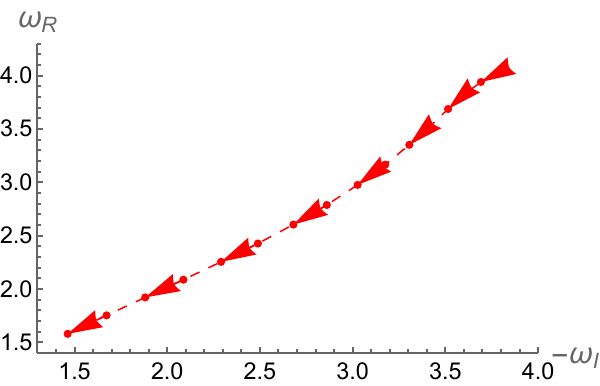}}
\hfill
\subfigure[$z=1.65$]{\label{sfm0kzb} 
\includegraphics[width=1.2in]{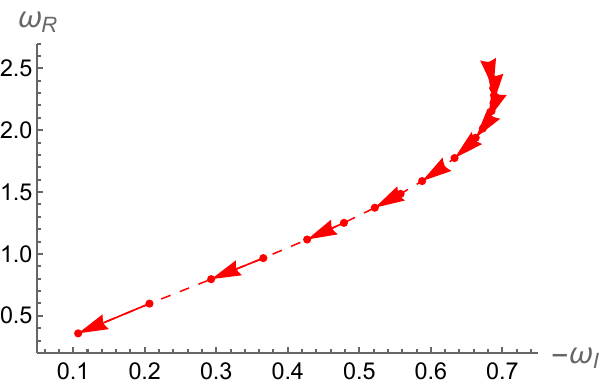}}
\hfill
\subfigure[$z=1.7$]{\label{sfm0kzc} 
\includegraphics[width=1.2in]{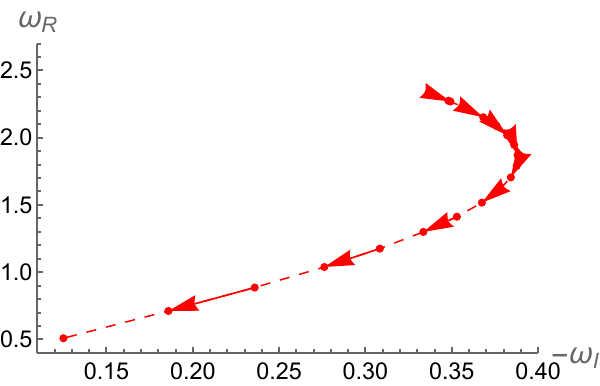}}
\hfill
\subfigure[$z=2$]{\label{sfm0kzd} 
\includegraphics[width=1.2in]{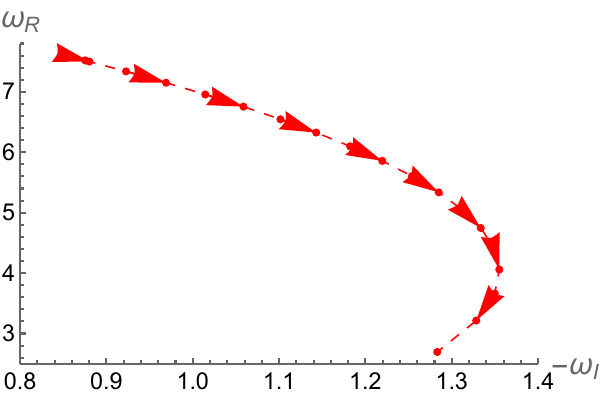}}
\hfill
\subfigure[$z=2.75$]{\label{sfm0kze} 
\includegraphics[width=1.2in]{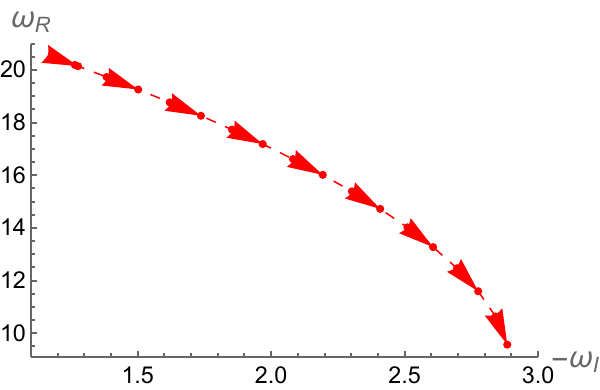}}
\hfill
\caption{$a=3,m_s=0,q_s=1,k=0,L=0,Q=1$, $b$ from 0 to near the extreme value. }\label{figkm0z}
\end{figure}

\begin{figure}[H]
\centering 
\subfigure[$z=1$]{\label{sfm1kza} 
\includegraphics[width=1.2in]{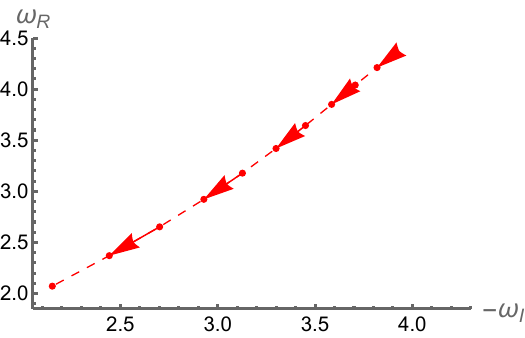}}
\hfill
\subfigure[$z=1.7$]{\label{sfm1kzb} 
\includegraphics[width=1.2in]{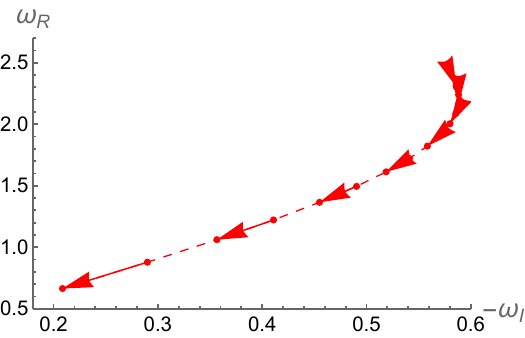}}
\hfill
\subfigure[$z=1.75$]{\label{sfm1kzc} 
\includegraphics[width=1.2in]{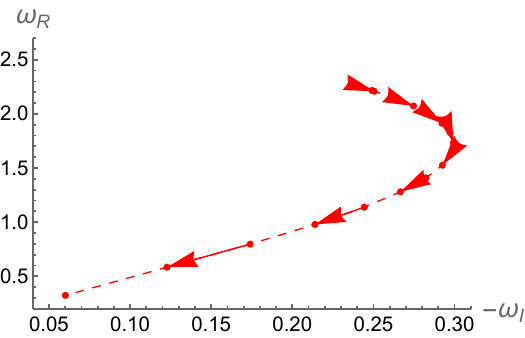}}
\hfill
\subfigure[$z=2$]{\label{sfm1kzd} 
\includegraphics[width=1.2in]{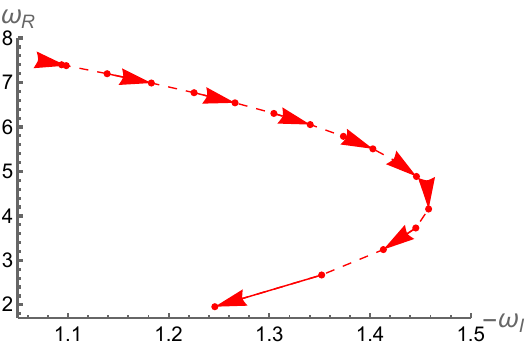}}
\hfill
\subfigure[$z=2.8$]{\label{sfm1kze} 
\includegraphics[width=1.2in]{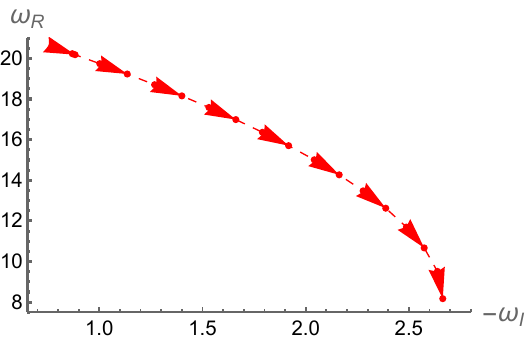}}
\hfill
\caption{$a=3,m_s=0,q_s=1,k=1,L=0,Q=1$, $b$ from 0 to near the extreme value.}\label{figkm1z}
\end{figure}

\begin{figure}[H]
\centering 
\subfigure[$z=1$]{\label{asfm1kza} 
\includegraphics[width=1.2in]{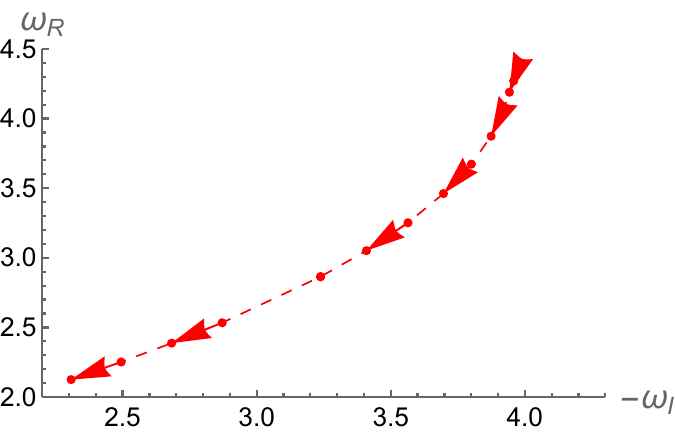}}
\hfill
\subfigure[$z=1.6$]{\label{asfm1kzb} 
\includegraphics[width=1.2in]{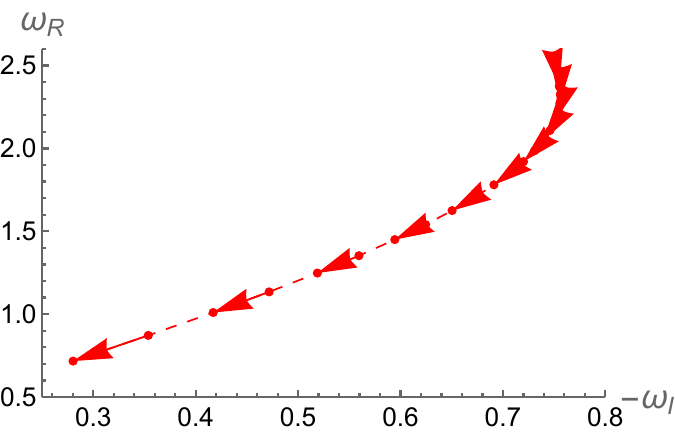}}
\hfill
\subfigure[$z=1.65$]{\label{asfm1kzc} 
\includegraphics[width=1.2in]{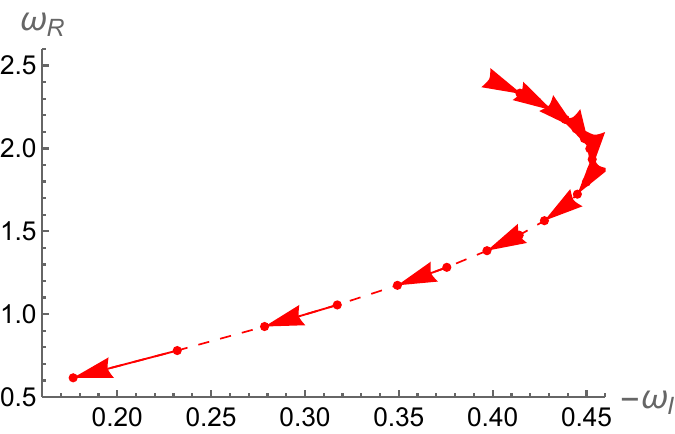}}
\hfill
\subfigure[$z=2$]{\label{asfm1kzd} 
\includegraphics[width=1.2in]{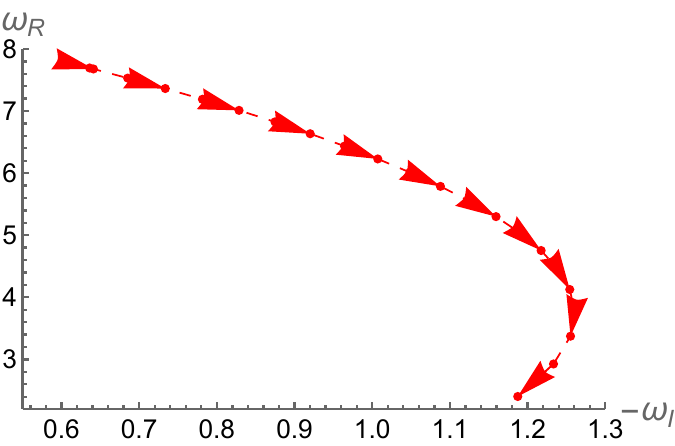}}
\hfill
\subfigure[$z=2.7$]{\label{asfm1kze} 
\includegraphics[width=1.2in]{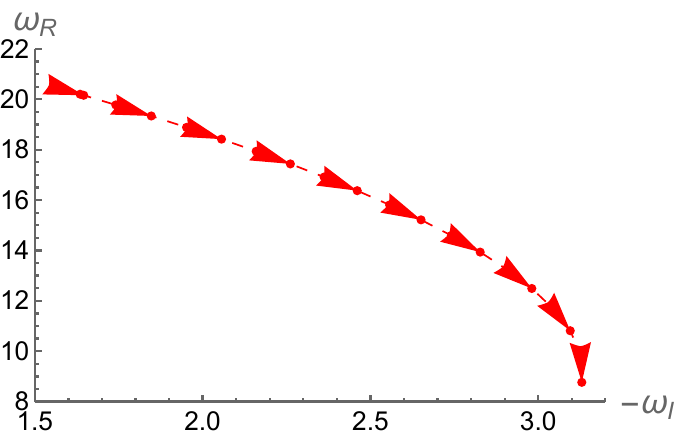}}
\hfill
\caption{$a=3,m_s=0,q_s=1,k=-1,L=0,Q=1$, $b$ from 0 to near the extreme value.}\label{figk-m1z}
\end{figure}

The Figs.\ref{figkm0z}-\ref{figk-m1z} displays the trajectories of quasinormal frequencies $\omega = \omega_R + i \omega_I$ in the complex frequency plane for a hairy Lifshitz black hole under perturbations of a charged, massless scalar field. The focus is on how the variation of black hole charge $b \in (0, b_{\text{ext}})$ affects the QNMs. The fixed parameters used include: black hole mass $a = 3$, scalar field mass $m_s = 0$, scalar field charge $q_s = 1$, spatial curvature parameter $k=0, \pm 1$, and $L = 0$ denoting the fundamental mode. The five subplots correspond to different values of the Lifshitz dynamical critical exponent $z$, ranging from $z = 1$ to $z = 2.75$. As the black hole charge $b$ increases (indicated by the direction of the arrows), the QNMs trace different trajectories in the complex plane, with their behavior significantly depending on the value of $z$.
\unskip
\unskip

When $z$ is small (e.g., $z = 1$ and $1.7$), both $\omega_R$ and $|\omega_I|$ decrease as $q$ increases, indicating a reduction in the oscillation frequency and a slower decay rate—that is, the perturbations last longer. For intermediate values such as $z = 1.75$ and $2$, $|\omega_I|$ shows a non-monotonic behavior: it first increases and then decreases, suggesting a shift in damping characteristics at a certain threshold. When $z = 2.8$, the trend is completely reversed—$|\omega_I|$ increases monotonically with $b$, implying that perturbations decay more rapidly as the black hole charge increases. These results indicate that the Lifshitz exponent $z$ plays a crucial role in determining the stability of the black hole, governing whether perturbations tend to persist or dissipate quickly.

\begin{figure}[H]
\centering 
\subfigure[$q_s=0.01$]{\label{msf0kqa} 
\includegraphics[width=1.2in]{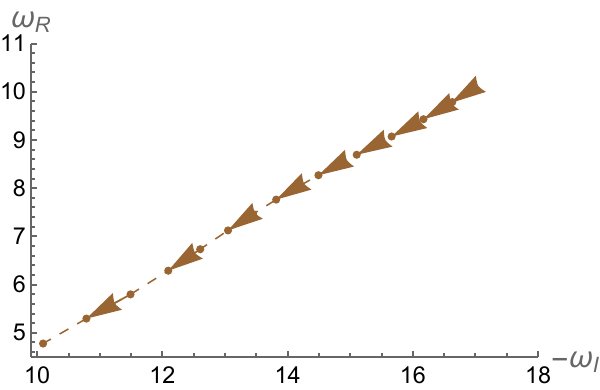}}
\hfill
\subfigure[$q_s=0.6$]{\label{msf0kqb} 
\includegraphics[width=1.2in]{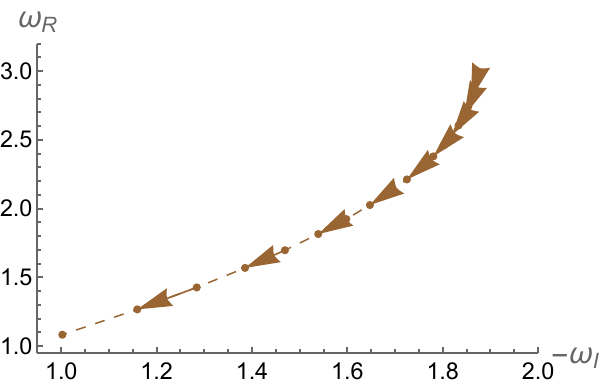}}
\hfill
\subfigure[$q_s=0.7$]{\label{msf0kqc} 
\includegraphics[width=1.2in]{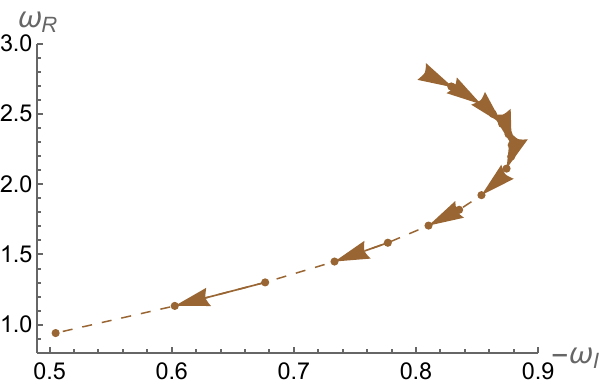}}
\hfill
\subfigure[$q_s=2$]{\label{msf0kqd} 
\includegraphics[width=1.2in]{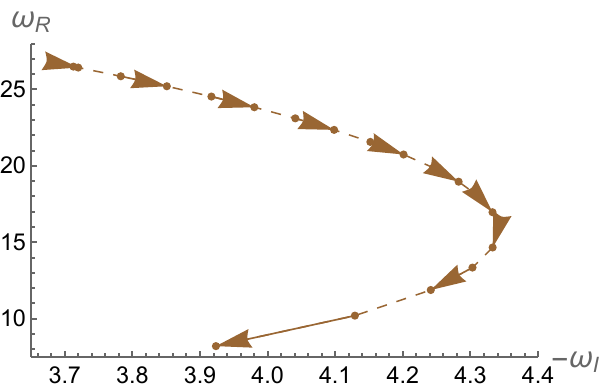}}
\hfill
\subfigure[$q_s=2.2$]{\label{msf0kqe} 
\includegraphics[width=1.2in]{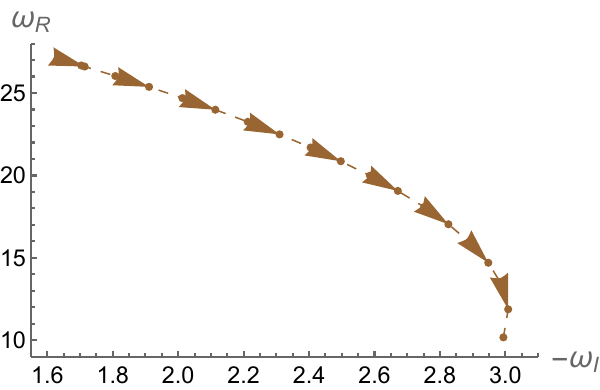}}
\hfill
\caption{$a=3,m_s=0,k=0,z=2,L=0,Q=1$, $b$ from 0 to near the extreme value.}\label{figk0qsm}
\end{figure}

\begin{figure}[H]
\centering 
\subfigure[$q_s=0.01$]{\label{msf1kqa} 
\includegraphics[width=1.2in]{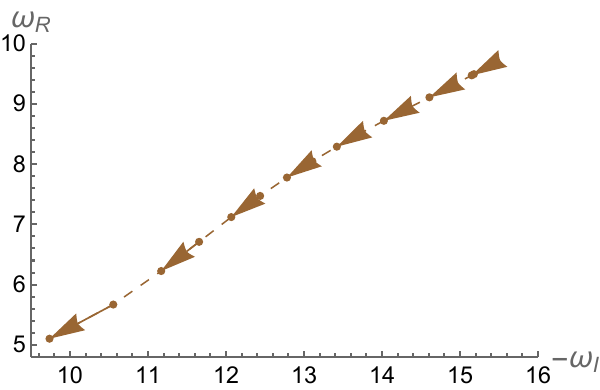}}
\hfill
\subfigure[$q_s=0.6$]{\label{msf1kqb} 
\includegraphics[width=1.2in]{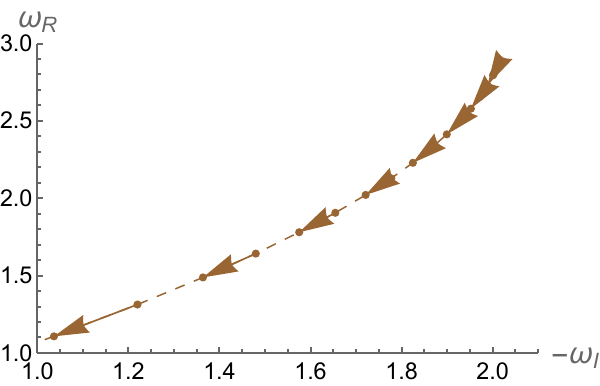}}
\hfill
\subfigure[$q_s=0.7$]{\label{msf1kqc} 
\includegraphics[width=1.2in]{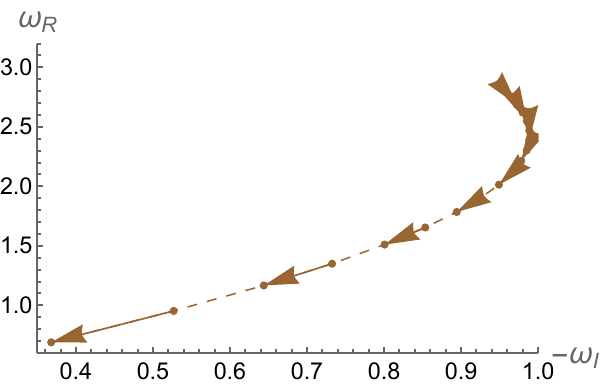}}
\hfill
\subfigure[$q_s=2$]{\label{msf1kqd} 
\includegraphics[width=1.2in]{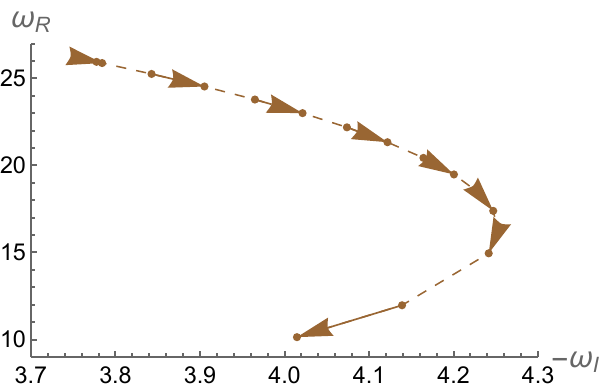}}
\hfill
\subfigure[$q_s=2.3$]{\label{msf1kqe} 
\includegraphics[width=1.2in]{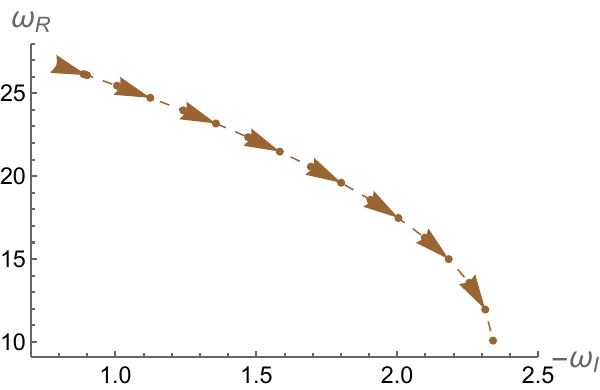}}
\hfill
\caption{$a=3,m_s=0,k=1,z=2,L=0,Q=1$, $b$ from 0 to near the extreme value.}\label{figkm1qs}
\end{figure}

\begin{figure}[H]
\centering 
\subfigure[$q_s=0.01$]{\label{-msf1kqa} 
\includegraphics[width=1.2in]{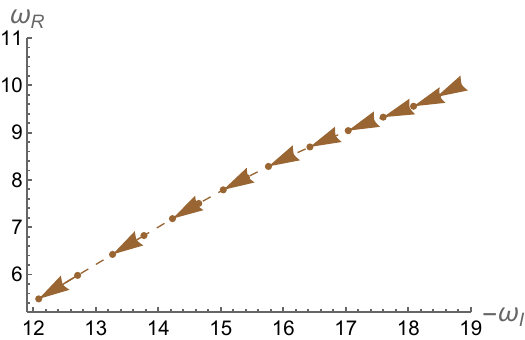}}
\hfill
\subfigure[$q_s=0.6$]{\label{-msf1kqb} 
\includegraphics[width=1.2in]{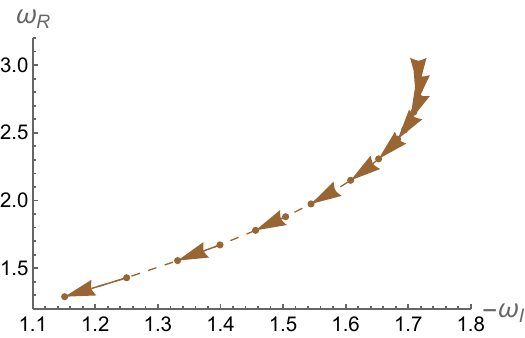}}
\hfill
\subfigure[$q_s=0.7$]{\label{-msf1kqc} 
\includegraphics[width=1.2in]{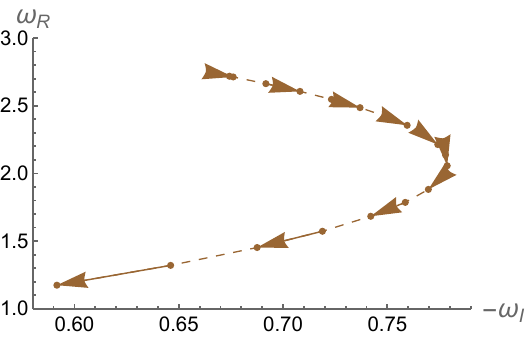}}
\hfill
\subfigure[$q_s=2$]{\label{-msf1kqd} 
\includegraphics[width=1.2in]{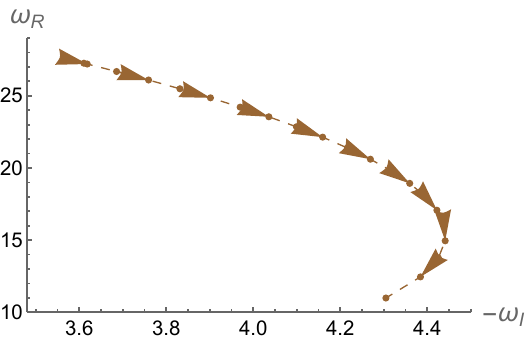}}
\hfill
\subfigure[$q_s=2.3$]{\label{-msf1kqe} 
\includegraphics[width=1.2in]{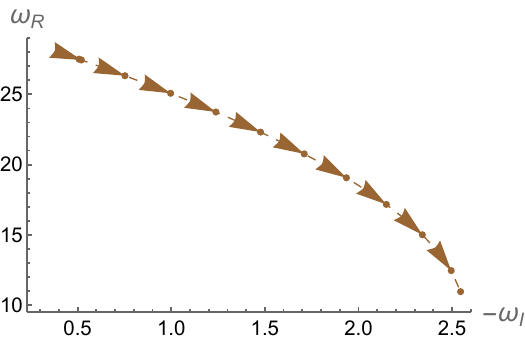}}
\hfill
\caption{$a=3,m_s=0,k=-1,z=2,L=0,Q=1$, $b$ from 0 to near the extreme value.}\label{-figkm1qs}
\end{figure}

The Figs.\ref{figk0qsm}-\ref{-figkm1qs} illustrates the trajectories of quasinormal frequencies $\omega = \omega_R + i \omega_I$ in the complex frequency plane for a hairy Lifshitz black hole under perturbations from a charged, massless scalar field. The focus is on the effect of the scalar field charge $q_s$ on the behavior of QNMs. The fixed parameters used are: black hole mass $a = 3$, scalar field mass $m_s = 0$, Lifshitz dynamical critical exponent $z = 2$, spatial curvature parameter $k=0, \pm 1$, and $L = 0$ indicating the fundamental mode. The five subfigures correspond to different values of the scalar field charge $q_s$. As the black hole charge $b$ increases (as indicated by the direction of the arrows), the QNMs trace distinct trajectories in the complex plane, and their behavior depends significantly on the value of $q_s$.
\unskip
\unskip

When $q_s$ is small (e.g., $q_s = 0.01$ and $0.6$), both $\omega_R$ and $|\omega_I|$ decrease with increasing $q$, indicating a lower oscillation frequency and a slower decay rate—in other words, the perturbations persist for a longer duration. At intermediate values such as $q_s = 0.7$ and $2$, $|\omega_I|$ exhibits a non-monotonic behavior: it first increases and then decreases, suggesting a transition in the damping characteristics at a certain threshold. When $q_s = 2.2$, the trend is fully reversed, and $|\omega_I|$ increases monotonically with $b$, implying that the perturbations decay faster as the black hole charge increases. These results demonstrate that the scalar field charge $q_s$ has a significant influence on the stability of the black hole, determining whether the perturbations tend to persist or dissipate rapidly.

\subsection{Massive Charged Scalar Perturbation}

We display here only the quasinormal modes for massive scalar perturbations with $k=0$, while the results for $k = \pm 1$ are included in Appendix A for reference.

The Fig.\ref{figk0z} displays the trajectories of the quasinormal frequencies $\omega = \omega_R + i \omega_I$ in the complex frequency plane for scalar perturbations with mass and charge in the background of charged Lifshitz black holes with scalar hair. The impact of varying the black hole charge $b $ from 0 to near the extreme value on the QNMs is examined. The fixed parameters are: black hole mass $a =3$, scalar field mass $m_s = 1$, scalar field charge $q_s = 1$, curvature parameter $k = 0$, and $L = 0$ indicating the fundamental mode. The five subfigures correspond to different values of the Lifshitz dynamical critical exponent $z$, ranging from $z = 1$ to $z = 2.75$. As the black hole charge $b$ increases (indicated by the direction of the arrows), the QNMs trace different trajectories in the complex plane, and their behavior is significantly influenced by the value of $z$.
\begin{figure}[H]
\centering 
\subfigure[$z=1$]{\label{sf0kza} 
\includegraphics[width=1.2in]{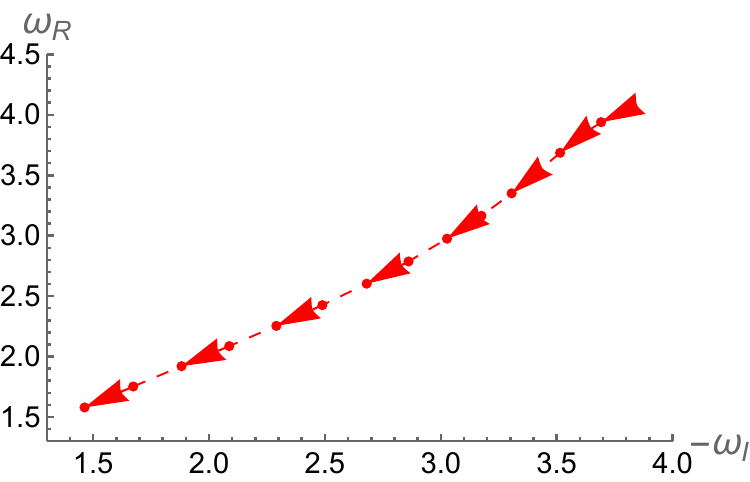}}
\hfill
\subfigure[$z=1.65$]{\label{sf0kzb} 
\includegraphics[width=1.2in]{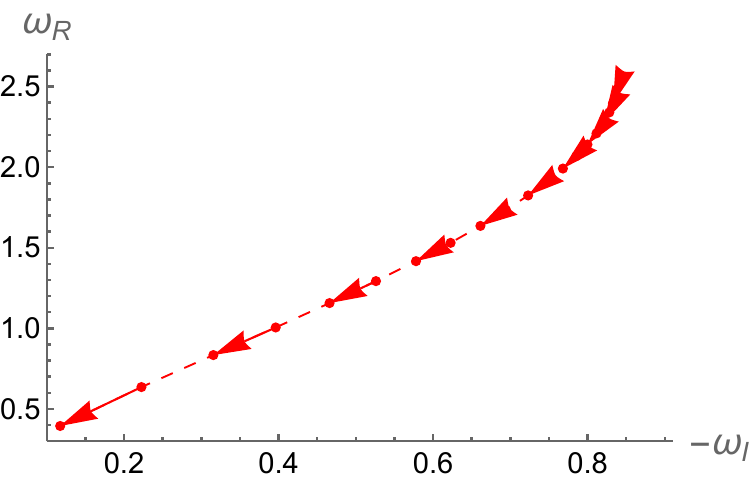}}
\hfill
\subfigure[$z=1.7$]{\label{sf0kzc} 
\includegraphics[width=1.2in]{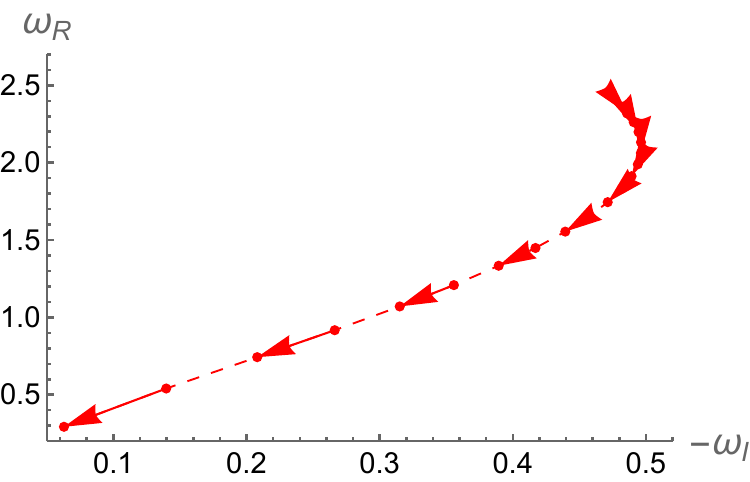}}
\hfill
\subfigure[$z=2$]{\label{sf0kzd} 
\includegraphics[width=1.2in]{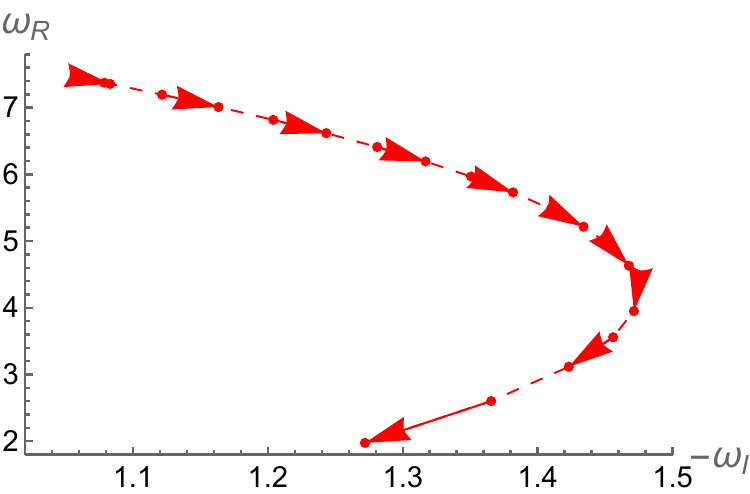}}
\hfill
\subfigure[$z=2.75$]{\label{sf0kze} 
\includegraphics[width=1.2in]{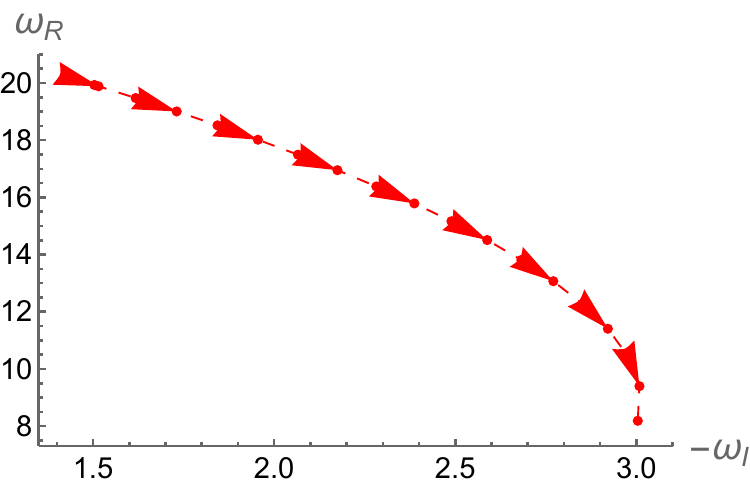}}
\hfill
\caption{$a=3,m_s=1,q_s=1,k=0,L=0,Q=1$, $b$ from 0 to near the extreme value.}\label{figk0z}
\end{figure}

For smaller values of $z$ (e.g., $z = 1$ and $1.7$), both $\omega_R$ and $|\omega_I|$ decrease as $q$ increases, indicating that the oscillation frequency of the perturbation decreases and the decay slows down, meaning the perturbation persists longer. For $z = 1.7$ and $z = 2$, $|\omega_I|$ exhibits non-monotonic behavior: it first increases and then decreases, suggesting a transition in the decay properties at a certain threshold of $q$. When $z = 2.75$, the trend completely reverses, and $|\omega_I|$ increases monotonically with $b$, implying that the perturbation decays faster as the black hole becomes more charged. These results indicate that the Lifshitz exponent $z$ has a significant influence on the stability of the black hole, determining whether the perturbation tends to persist for a long time or decay rapidly.

\begin{figure}[H]
\centering 
\subfigure[$q_s=0.01$]{\label{sf0kqa} 
\includegraphics[width=1.2in]{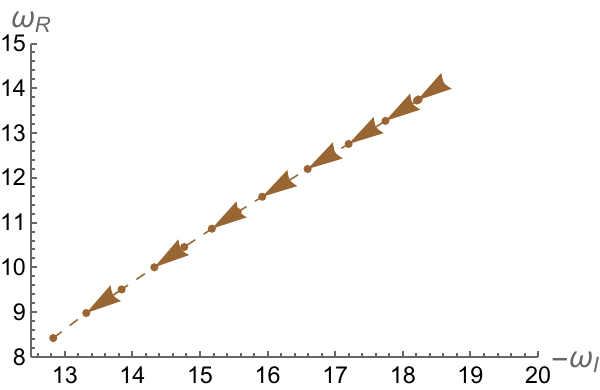}}
\hfill
\subfigure[$q_s=0.6$]{\label{sf0kqb} 
\includegraphics[width=1.2in]{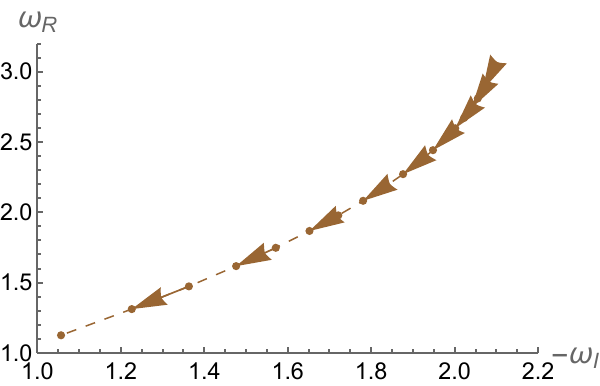}}
\hfill
\subfigure[$q_s=0.7$]{\label{sf0kqc} 
\includegraphics[width=1.2in]{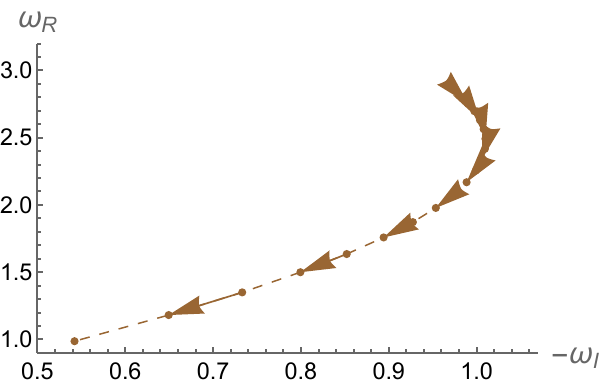}}
\hfill
\subfigure[$q_s=2$]{\label{sf0kqd} 
\includegraphics[width=1.2in]{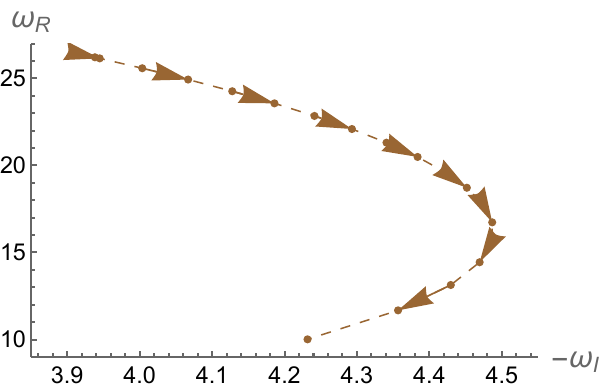}}
\hfill
\subfigure[$q_s=2.2$]{\label{sf0kqe} 
\includegraphics[width=1.2in]{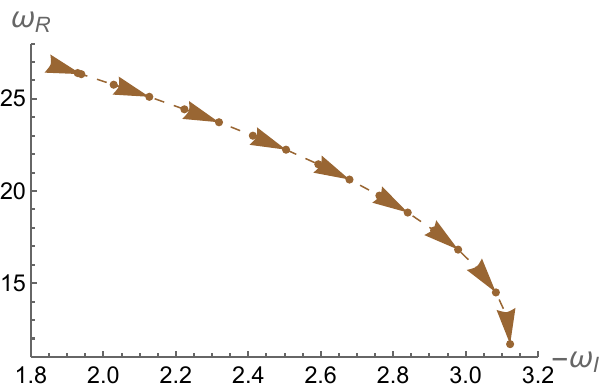}}
\hfill
\caption{$a=3,m_s=1,k=0,z=2,L=0,Q=1$, $b$ from 0 to near the extreme value.}\label{figk0qs}
\end{figure}

The Fig.\ref{figk0qs} show the trajectories of quasinormal frequencies $\omega = \omega_R + i \omega_I$ in the complex frequency plane for a hairy Lifshitz black hole under perturbations of a charged, massive scalar field. The main focus is on how the scalar field charge $q_s$ affects the behavior of QNMs. The fixed parameters used are: black hole mass $a = 3$, scalar field mass $m_s = 1$, Lifshitz dynamical critical exponent $z = 2$, spatial curvature parameter $k = 0$, and $L = 0$ denoting the fundamental mode. The five subplots correspond to different values of the scalar field charge $q_s$. As the black hole charge $b$ increases (indicated by the direction of the arrows), the QNMs trace out different paths in the complex plane, with their behavior strongly depending on the value of $q_s$.

When $q_s$ is small (e.g., $q_s = 0.01$ and $0.6$), both $\omega_R$ and $|\omega_I|$ decrease with increasing $b$, indicating a lower oscillation frequency and slower decay rate—i.e., the perturbations persist longer. For $q_s = 0.7$ and $2$, $|\omega_I|$ exhibits a non-monotonic trend, first increasing and then decreasing, suggesting a transition in the damping behavior at a certain threshold. When $q_s = 2.2$, the trend is completely reversed: $|\omega_I|$ increases monotonically with $b$, implying that the perturbations decay faster as the black hole charge increases. These results demonstrate that the scalar field charge $q_s$ significantly influences black hole stability, determining whether perturbations tend to linger or dissipate rapidly.

\subsection{$m_s-$ Dependence}

\begin{figure}[H]
\centering 
\subfigure[$m_s=0.01$]{\label{sf0kma} 
\includegraphics[width=1.5in]{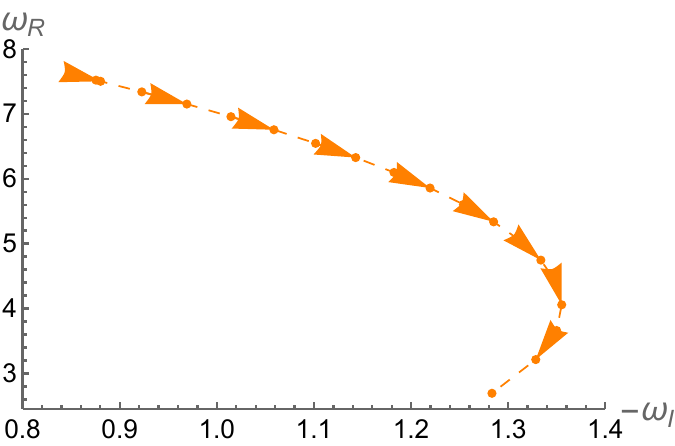}}
\hfill
\subfigure[$m_s=3$]{\label{sf0kmb} 
\includegraphics[width=1.5in]{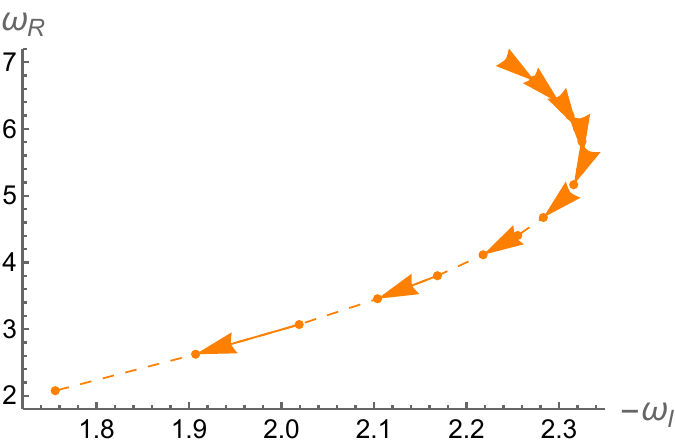}}
\hfill
\subfigure[$m_s=3.8$]{\label{sf0kmc} 
\includegraphics[width=1.5in]{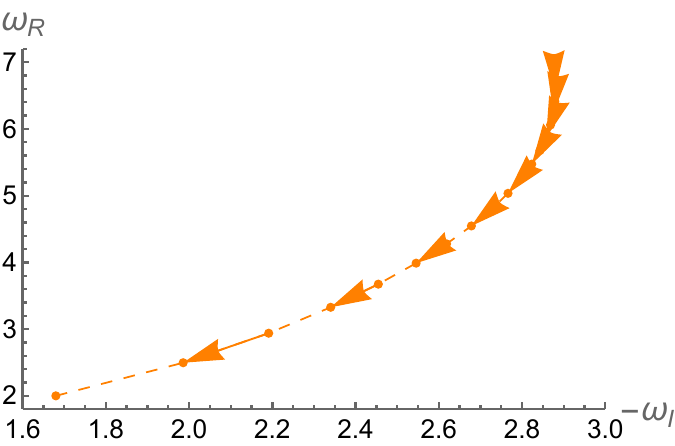}}
\hfill
\subfigure[$m_s=6$]{\label{sf0kmd} 
\includegraphics[width=1.5in]{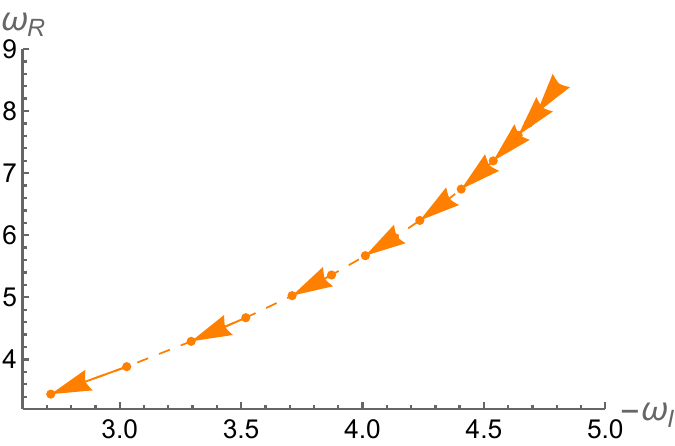}}
\hfill
\caption{$a=3,q_s=1,k=0,z=2,L=0,Q=1$, $b$ from 0 to near the extreme value.}\label{figk0ms}
\end{figure}

The Fig.\ref{figk0ms} illustrate the trajectories of quasinormal frequencies $\omega = \omega_R + i\omega_I$ in the complex frequency plane for a hairy Lifshitz black hole under perturbations of a charged, massive scalar field. The focus is on how the scalar field mass $m_s$ affects the behavior of QNMs. Fixed parameters include the black hole mass $a = 3$, scalar field charge $q_s = 1$, Lifshitz dynamical critical exponent $z = 2$, and $L = 0$, representing the fundamental mode. The black hole charge is set to $Q = 1$, and the scalar perturbation charge $b$ varies within the range from 0 to near the extreme value. The arrows indicate the direction of increasing $b$. Each subplot corresponds to a different value of the scalar field mass: $m_s = 0.01, 3, 3.5, 6$.

For spatial curvature parameters $k = 0$, when the scalar mass $m_s$ is small (Fig. (a)), the QNM curves exhibit decreasing $\omega_R$ and $|\omega_I|$ with increasing $b$, indicating a lower oscillation frequency and a slower decay rate, i.e., a longer-lasting perturbation. As $m_s$ increases, $|\omega_I|$ exhibits a non-monotonic behavior—first increasing and then decreasing—suggesting a transition in the damping characteristics at a certain threshold.
In contrast, when the spatial curvature is $k = 1$, non-monotonic features already appear in Fig. (a) even at small $m_s$: as $b$ increases, $\omega_R$ decreases monotonically, while $|\omega_I|$ first increases and then decreases, indicating a reversal in the decay rate in the intermediate $b$ region. As $m_s$ increases (Figs. (b)-(d)), this non-monotonicity in $|\omega_I|$ gradually disappears. When $m_s = 6$, regardless of the value of $k$, $|\omega_I|$ decreases monotonically with increasing $b$, implying that a larger scalar mass enhances the stability of the perturbation and prolongs the decay time. Overall, the scalar field mass is a key factor in determining the behavior of perturbations and the dynamical stability of the black hole.

\section{Conclusion and Discussion}
\label{sec:level6}

In this work, we have investigated the quasinormal modes (QNMs) of charged scalar perturbations around four-dimensional Lifshitz black holes with scalar hair, focusing on the interplay between the Lifshitz exponent, scalar field parameters, and black hole charge in determining the stability and dynamical properties of the system. Both massless and massive scalar fields were considered, and their effects were systematically studied under varying spatial curvature ($k=0, \pm1$) and black hole charge $b$. Our findings highlight a rich landscape of dynamical behavior that depends sensitively on the values of key parameters such as the Lifshitz dynamical critical exponent $z$, scalar field charge $q_s$, and scalar field mass $m_s$.

For massless charged scalar fields, we found that the Lifshitz exponent $z$ plays a decisive role in modulating the evolution of perturbations. When $z$ is small , both the real part $\omega_R$ and the magnitude of the imaginary part $|\omega_I|$ decrease with increasing black hole charge $b$, indicating that the perturbations become less oscillatory and decay more slowly, hence persisting longer. However, for moderate values of $z$ , a non-monotonic behavior in $|\omega_I|$ emerges—first increasing and then decreasing with $b$, suggesting a critical threshold beyond which the damping mechanism reverses. At higher values like $z=2.8$, the trend completely reverses: $|\omega_I|$ increases monotonically with $b$, leading to faster decay and improved stability. These results emphasize the crucial role of the Lifshitz exponent in controlling the timescale of perturbation dissipation.

When we varied the scalar field charge $q_s$ while keeping the field massless and fixing $z=2$, a similar nontrivial pattern emerged. For small $q_s$, the system exhibits reduced oscillation and longer-lived perturbations with increasing $b$. As $q_s$ increases to intermediate values, the decay rate shows a non-monotonic trend. For large charges such as $q_s=2.2$, perturbations decay faster with increasing black hole charge. This again signals a transition in dynamical behavior and implies that scalar charge is a critical parameter governing the system’s response to perturbations.

For massive charged scalar fields, the qualitative picture remains but becomes more intricate due to the additional parameter $m_s$. Our analysis reveals that increasing scalar mass tends to suppress oscillations and prolong perturbation decay, especially when $k = 0$ or $k = -1$. A notable non-monotonic transition in the imaginary part of QNMs arises for intermediate scalar masses, which then vanishes as $m_s$ becomes large—suggesting enhanced damping stability in such regimes. Interestingly, in the case of positive curvature ($k = 1$), this non-monotonicity in decay rate appears even at small $m_s$, highlighting the influence of spatial geometry on perturbative dynamics. When the scalar field mass is large ($m_s=6$), the decay rate decreases monotonically with $b$ regardless of the curvature, indicating that a sufficiently massive scalar field can enhance the black hole’s resistance to dynamical perturbations.

In conclusion, our results show that the stability and decay characteristics of scalar perturbations in Lifshitz black holes with scalar hair are governed by a subtle competition among the black hole charge, Lifshitz exponent, scalar field charge, and scalar mass. The appearance of non-monotonic behaviors in the damping rates underlines the possibility of critical transitions in the perturbative dynamics. This study provides not only a deeper understanding of the stability of Lifshitz black holes but also serves as a starting point for exploring more general dynamical processes in non-AdS, anisotropic geometries. Future work may extend this analysis to vector or gravitational perturbations, or explore the holographic implications of QNMs in strongly coupled field theories with Lifshitz scaling.

\vspace{1cm} 
{\bf Acknowledgments}
\vspace{1cm}

We gratefully acknowledge support by the Natural Science Foundation of China (NNSFC) (Grant No.12365009), Jiangxi Provincial Natural Science Foundation (Grant No. 20232BAB201039) and Natural Science Basic Research Program of Shaanxi (Program No.2023-JC-QN-0053).

\appendix
\section[\appendixname~\thesection]{Supplementary Plots for Massive Scalar Field Perturbations}

Here we include the QNM spectra for massive scalar field perturbations with $k=+1$ and $k=-1$, which were omitted from the main text for brevity. The results are qualitatively similar to those for $k=0$, and are presented here for completeness.
\begin{figure}[H]
\centering
\subfigure[$z=1.2$]{\label{sf1kza} 
\includegraphics[width=1.2in]{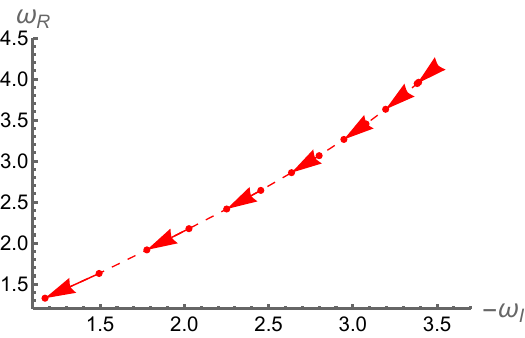}}
\hfill
\subfigure[$z=1.7$]{\label{sf1kzb} 
\includegraphics[width=1.2in]{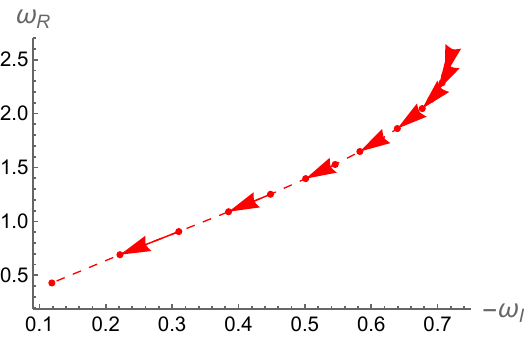}}
\hfill
\subfigure[$z=1.75$]{\label{sf1kzc} 
\includegraphics[width=1.2in]{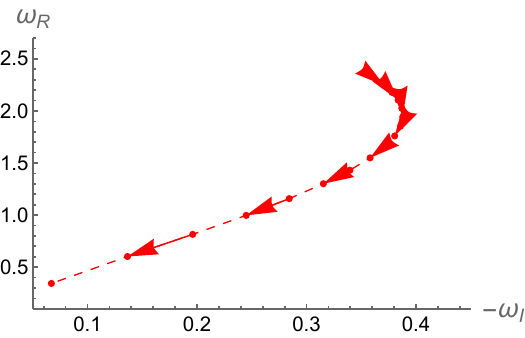}}
\hfill
\subfigure[$z=2$]{\label{sf1kzd} 
\includegraphics[width=1.2in]{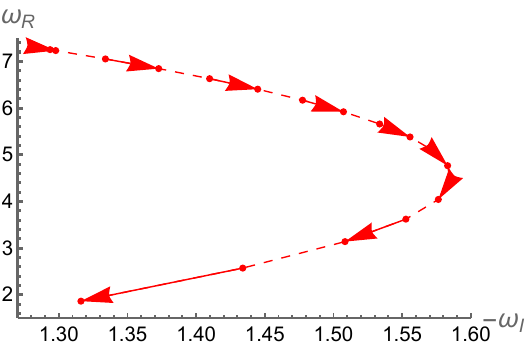}}
\hfill
\subfigure[$z=2.8$]{\label{sf1kze} 
\includegraphics[width=1.2in]{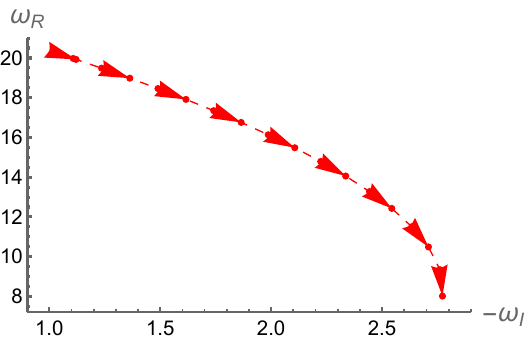}}
\hfill
\caption{$a=3,m_s=1,q_s=1,k=1,L=0,Q=1$, $b$ from 0 to near the extreme value.}\label{figk1z}
\end{figure}

\begin{figure}[H]
\centering
\subfigure[$z=1.2$]{\label{sf2kza} 
\includegraphics[width=1.2in]{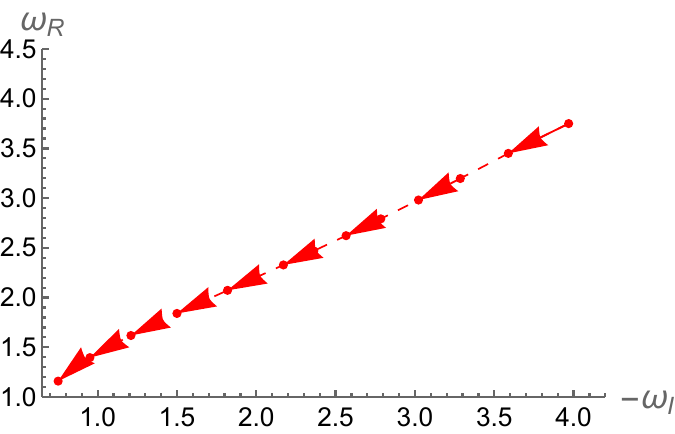}}
\hfill
\subfigure[$z=1.6$]{\label{sf2kzb} 
\includegraphics[width=1.2in]{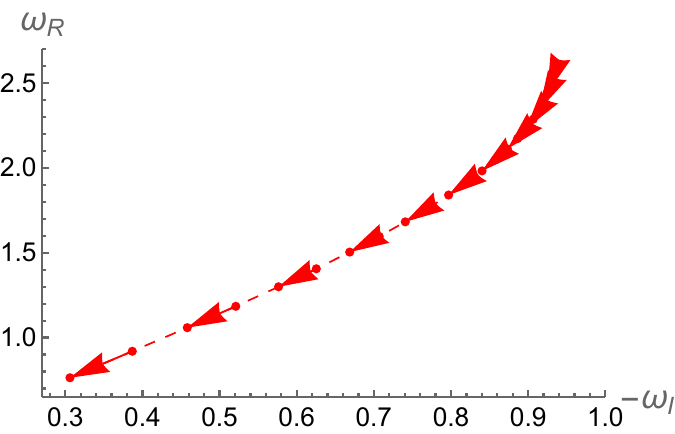}}
\hfill
\subfigure[$z=1.65$]{\label{sf2kzc} 
\includegraphics[width=1.2in]{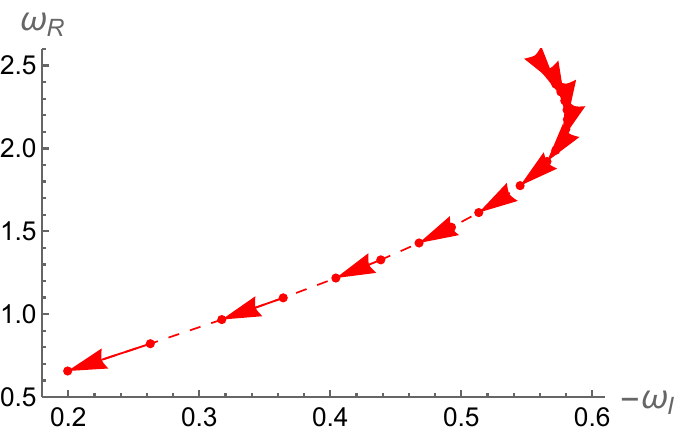}}
\hfill
\subfigure[$z=2$]{\label{sf2kzd} 
\includegraphics[width=1.2in]{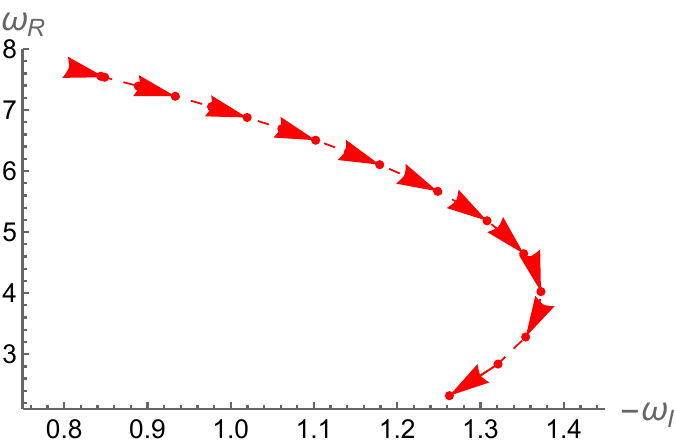}}
\hfill
\subfigure[$z=2.7$]{\label{sf2kze} 
\includegraphics[width=1.2in]{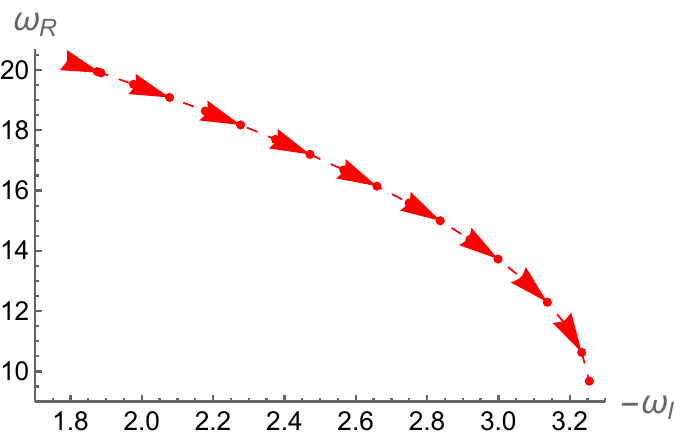}}
\hfill
\caption{$a=3,m_s=1,q_s=1,k=-1,L=0,Q=1$, $b$ from 0 to near the extreme value.}\label{figk2z}
\end{figure}

\begin{figure}[H]
\centering
\subfigure[$q_s=0.01$]{\label{sf1kqa} 
\includegraphics[width=1.2in]{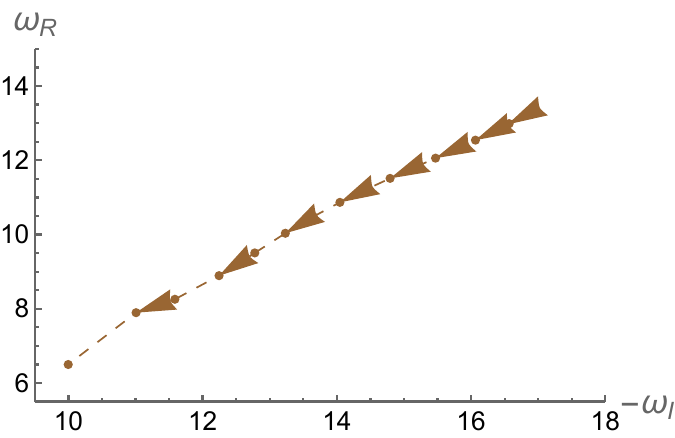}}
\hfill
\subfigure[$q_s=0.6$]{\label{sf1kqb} 
\includegraphics[width=1.2in]{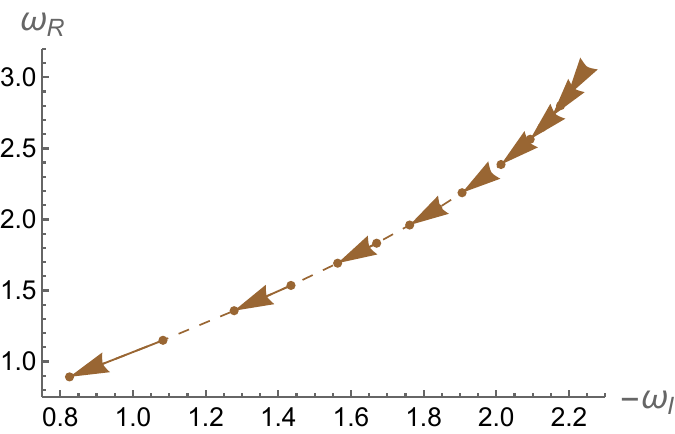}}
\hfill
\subfigure[$q_s=0.7$]{\label{sf1kqc} 
\includegraphics[width=1.2in]{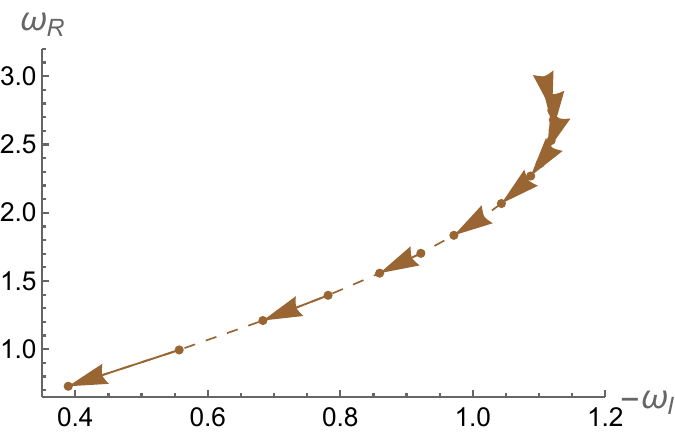}}
\hfill
\subfigure[$q_s=2$]{\label{sf1kqd} 
\includegraphics[width=1.2in]{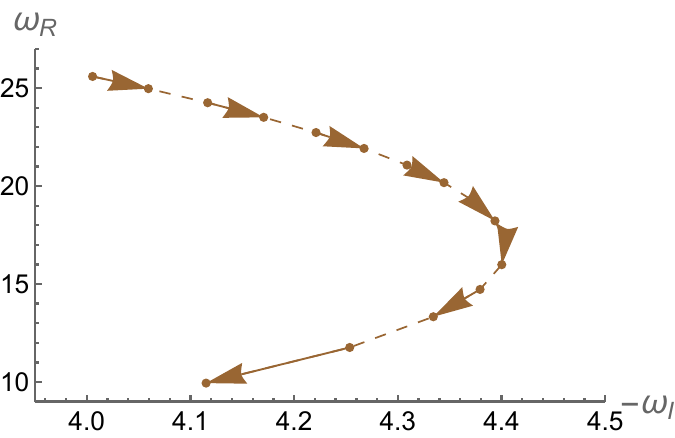}}
\hfill
\subfigure[$q_s=2.5$]{\label{sf1kqe} 
\includegraphics[width=1.2in]{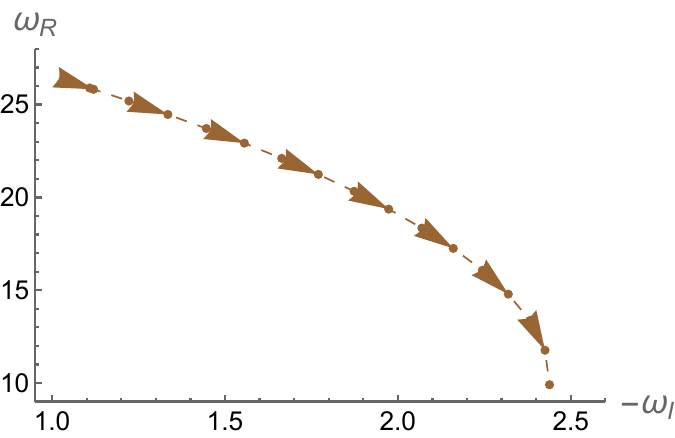}}
\hfill
\caption{$a=3,m_s=1,k=1,z=2,L=0,Q=1$, $b$ from 0 to near the extreme value.}\label{figk1qs}
\end{figure}

\begin{figure}[H]
\centering
\subfigure[$q_s=0.01$]{\label{sf2kqa} 
\includegraphics[width=1.2in]{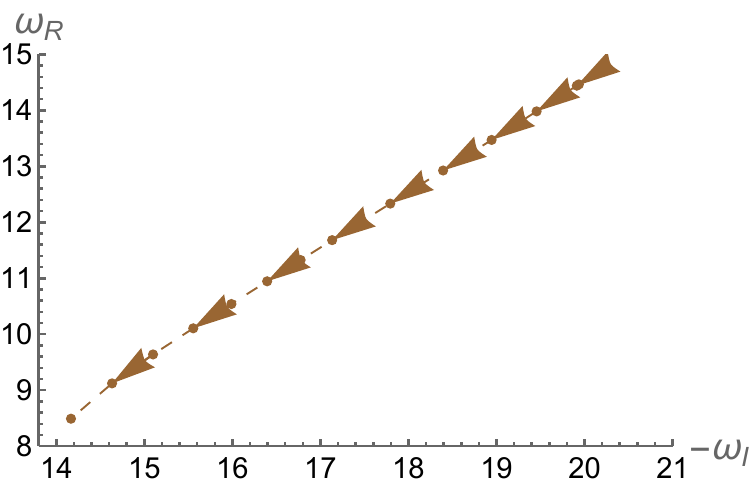}}
\hfill
\subfigure[$q_s=0.6$]{\label{sf2kqb} 
\includegraphics[width=1.2in]{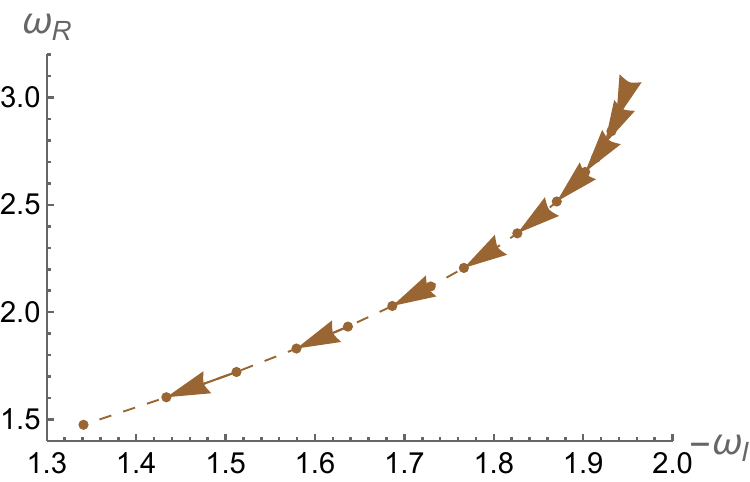}}
\hfill
\subfigure[$q_s=0.7$]{\label{sf2kqc} 
\includegraphics[width=1.2in]{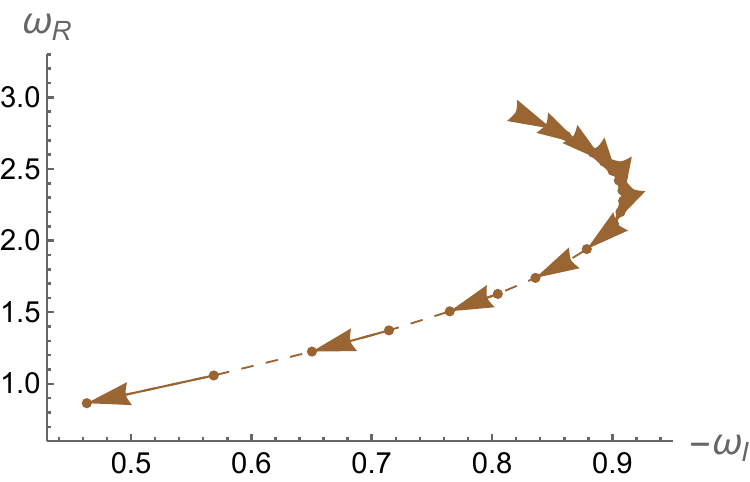}}
\hfill
\subfigure[$q_s=2$]{\label{sf2kqd} 
\includegraphics[width=1.2in]{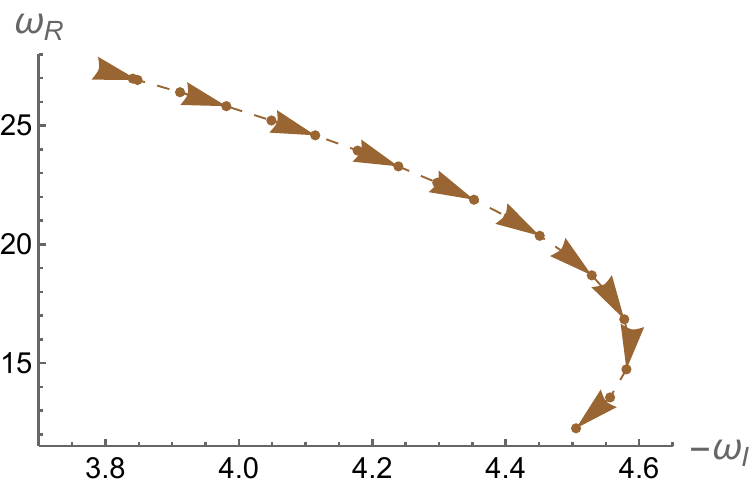}}
\hfill
\subfigure[$q_s=2.3$]{\label{sf2kqe} 
\includegraphics[width=1.2in]{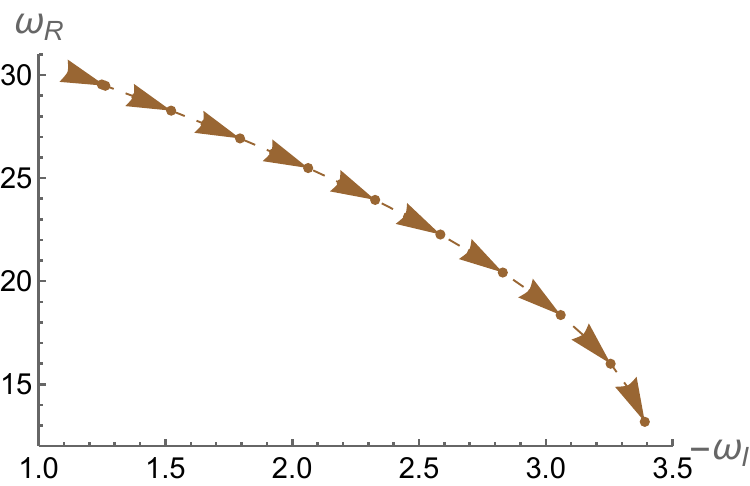}}
\hfill
\caption{$a=3,m_s=1,k=-1,z=2,L=0,Q=1$, $b$ from 0 to near the extreme value.}\label{figk2qs}
\end{figure}

\begin{figure}[H]
\centering
\subfigure[$m_s=0.01$]{\label{sf1kma} 
\includegraphics[width=1.5in]{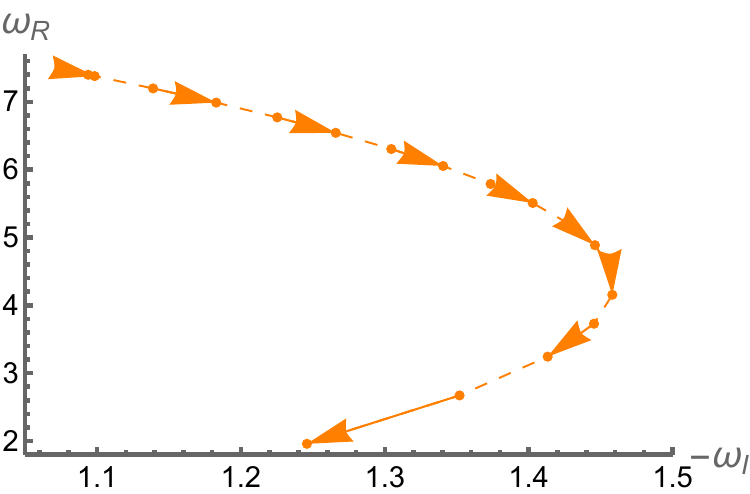}}
\hfill
\subfigure[$m_s=3$]{\label{sf1kmb} 
\includegraphics[width=1.5in]{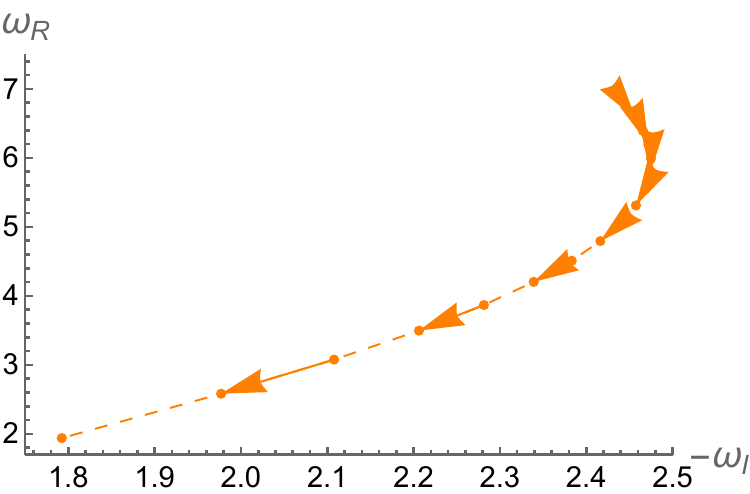}}
\hfill
\subfigure[$m_s=3.5$]{\label{sf1kmc} 
\includegraphics[width=1.5in]{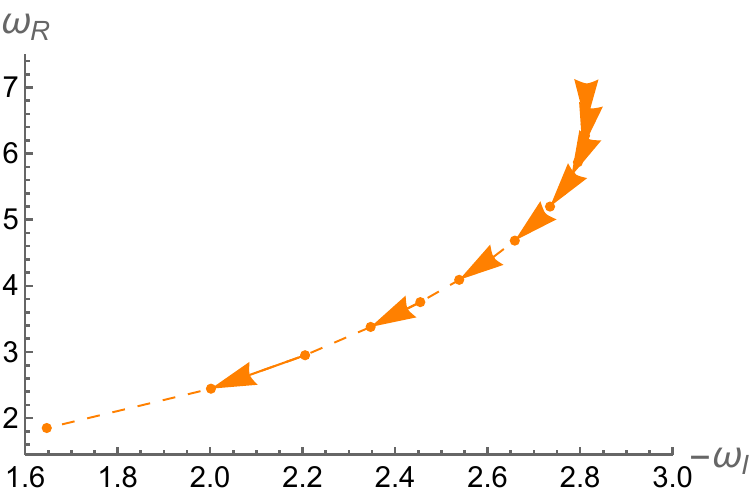}}
\hfill
\subfigure[$m_s=6$]{\label{sf1kmd} 
\includegraphics[width=1.5in]{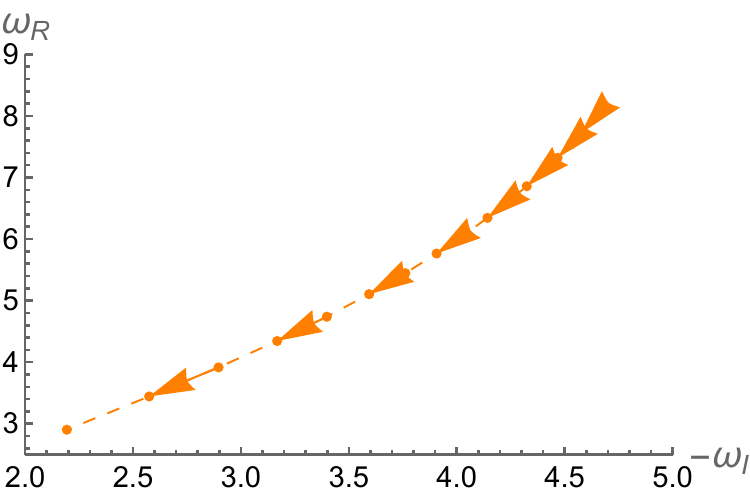}}
\hfill
\caption{$a=3,q_s=1,k=1,z=2,L=0,Q=1$, $b$ from 0 to near the extreme value.}\label{figk1ms}
\end{figure}

\begin{figure}[H]
\centering
\subfigure[$m_s=0.01$]{\label{sf2kma} 
\includegraphics[width=1.5in]{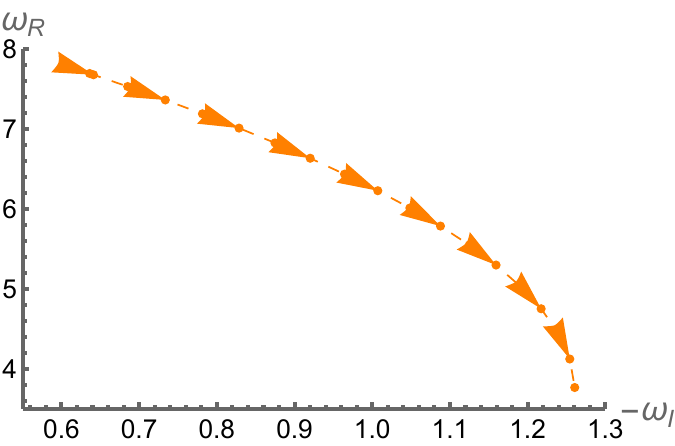}}
\hfill
\subfigure[$m_s=3$]{\label{sf2kmb} 
\includegraphics[width=1.5in]{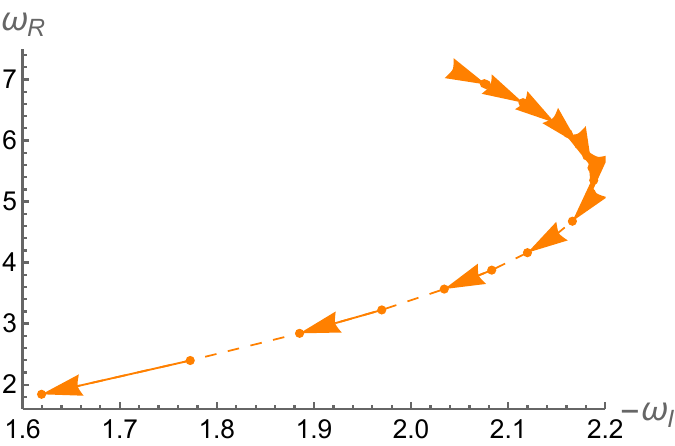}}
\hfill
\subfigure[$m_s=4.2$]{\label{sf2kmc} 
\includegraphics[width=1.5in]{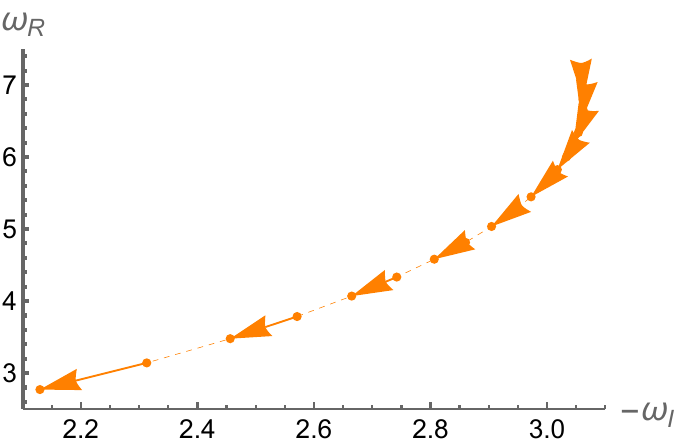}}
\hfill
\subfigure[$m_s=6$]{\label{sf2kmd} 
\includegraphics[width=1.5in]{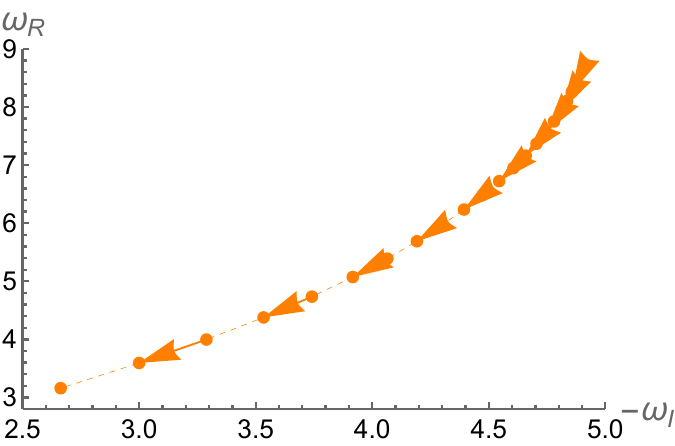}}
\hfill
\caption{$a=3,q_s=1,k=-1,z=2,L=0,Q=1$, $b$ from 0 to near the extreme value.}\label{figk2ms}
\end{figure}

\end{document}